\documentclass[lettersize,journal]{IEEEtran}
\usepackage{amsmath,amsfonts}
\usepackage{array}
\usepackage{textcomp}
\usepackage{stfloats}
\usepackage{url}
\usepackage{verbatim}
\usepackage{graphicx}
\usepackage{cite}
\usepackage{charlesmacros}
\usepackage{hyperref}
\usepackage{cleveref}
\usepackage{orcidlink}
\usepackage{tcolorbox}
\usepackage{footnote}
\usepackage{xr}
\BeforeBeginEnvironment{assbox}{\savenotes}
\AfterEndEnvironment{assbox}{\spewnotes}
\newtcolorbox{theobox}{colback=green!10, arc=3mm,colframe=red!0, boxrule=0pt}
\newtcolorbox{propbox}{colback=blue!5, arc=3mm,colframe=red!0, boxrule=0pt}
\newtcolorbox{lembox}{colback=blue!5, arc=3mm,colframe=red!0, boxrule=0pt}
\newtcolorbox{assbox}{colback=red!5, arc=3mm,colframe=red!0, boxrule=0pt}

\usepackage[caption=false,font=footnotesize]{subfig}
\usepackage[font=small,labelfont=bf]{caption}
\SetAlCapNameFnt{\footnotesize}
\SetAlCapFnt{\footnotesize}

\newcommand{\lA}{\widetilde{{A}}}
\newcommand{\cy}{\widehat{\bm{y}}}
\newcommand{\A}{{A}}
\newcommand{\cA}{\widehat{{A}}}
\newcommand{\cB}{\widehat{{B}}}
\newcommand{\Al}{\widehat{{A}}}

\newcommand{\y}{\bm{y}}
\newcommand{\ly}{\widetilde{\bm{y}}}
\newcommand{\X}{{X}}

\newcommand{\lX}{\widetilde{{X}}}
\newcommand{\z}{\bm{z}}
\newcommand{\B}{{B}}

\newcommand{\C}{{D}}
\newcommand{\cC}{\widehat{D}}
\newcommand{\cR}{\widehat{R}}

\newcommand{\ccr}{\widehat{r}}

\newcommand{\lB}{\widetilde{{B}}}

\newcommand{\accX}{\bar{\X}}
\DeclareMathOperator{\nullsp}{null}

\usetikzlibrary{hobby,backgrounds,calc}
\pgfdeclarelayer{background}
\pgfsetlayers{background,main}

\newcommand{\hobbyconvexpath}[2]{
[   
    create hobbyhullnodes/.code={
        \global\edef\namelist{#1}
        \foreach [count=\counter] \nodename in \namelist {
            \global\edef\numberofnodes{\counter}
            \node at (\nodename)
[draw=none,name=hobbyhullnode\counter] {};
        }
        \node at (hobbyhullnode\numberofnodes)
[name=hobbyhullnode0,draw=none] {};
        \pgfmathtruncatemacro\lastnumber{\numberofnodes+1}
        \node at (hobbyhullnode1)
[name=hobbyhullnode\lastnumber,draw=none] {};
    },
    create hobbyhullnodes
]
($(hobbyhullnode1)!#2!-90:(hobbyhullnode0)$)
\pgfextra{
  \gdef\hullpath{}
\foreach [
    evaluate=\currentnode as \previousnode using int(\currentnode-1),
    evaluate=\currentnode as \nextnode using int(\currentnode+1)
    ] \currentnode in {1,...,\numberofnodes} {
    \ifnum\currentnode=1\relax
    \xdef\hullpath{([closed=true]$(hobbyhullnode\currentnode)!#2!180:(hobbyhullnode\previousnode)$)
  ..($(hobbyhullnode\nextnode)!0.5!(hobbyhullnode\currentnode)$)}
    \else
    \xdef\hullpath{\hullpath
  ..($(hobbyhullnode\currentnode)!#2!180:(hobbyhullnode\previousnode)$)
  ..($(hobbyhullnode\nextnode)!0.5!(hobbyhullnode\currentnode)$)}
    \fi
    \ifx\currentnode\numberofnodes
    \else
    \xdef\hullpath{\hullpath
  ..($(hobbyhullnode\nextnode)!#2!-90:(hobbyhullnode\currentnode)$)}
    \fi
}
}
\hullpath
}

\begin{document}
\setlength\abovedisplayskip{2pt}
\setlength\belowdisplayskip{2pt}
\setlength\textfloatsep{2pt}

\title{A Distributed Adaptive Algorithm for Non-Smooth Spatial Filtering
Problems in Wireless Sensor Networks}

\author{
Charles Hovine \orcidlink{0000-0001-6657-1066}, Alexander Bertrand
\orcidlink{0000-0002-4827-8568}

\thanks{This project has received funding from the European Research Council
(ERC) under the European Union's Horizon 2020 research and innovation programme
(grant agreement No. 802895 and No. 101138304) and from the Flemish Government under the
``Onderzoeksprogramma Artifici\"ele Intelligentie (AI) Vlaanderen'' programme. Views and opinions expressed are however those of the author(s) only and do not necessarily reflect those of the European Union or ERC. Neither the European Union nor the ERC can be held responsible for them.}
\thanks{Charles Hovine and Alexander Bertrand are with the STADIUS Center for
        Dynamical Systems, Signal Processing and Data Analytics and with the Leuven.AI institute for Artificial Intelligence at KU Leuven,
        Leuven 3001, Belgium (e-mails: \{charles.hovine,
alexander.bertrand\}@esat.kuleuven.be).}

\thanks{A conference precursor of this manuscript has been published in \cite{hovine2023distributed}.}}



\maketitle

%\wip{ 
%        Key messages:
%        \begin{enumerate}
%                \item NS-DASF vs. other approaches
%                \item Scope of NS-DASF
%                \item Algorithmic flow of NS-DASF
%                \item NS-DASF converges
%                \item NS-DASF can be used for node selection
%        \end{enumerate}
%        Target audience: Statistical signal processing practitionners, with knowledge of smooth multivariate optimization.
%}

\begin{abstract}
        A wireless sensor network often relies on a fusion center to process the data
collected by each of its sensing nodes. Such an approach relies on the continuous
transmission of raw data to the fusion center, which typically has a major impact on
the sensors' battery life.
To address this issue in
the particular 
context of spatial filtering and signal fusion problems, we recently proposed the Distributed
Adaptive Signal Fusion (DASF) algorithm, which distributively computes a spatial
filter expressed as the solution of a smooth optimization problem involving the
network-wide sensor signal statistics. In this work,
we show that the DASF algorithm can be extended to compute the filters associated
with  a certain class of non-smooth optimization problems. This extension makes
the addition of sparsity-inducing norms to the problem's cost function possible, allowing sensor selection to be performed in
a distributed fashion, alongside the filtering task of interest, thereby further reducing the network's energy consumption. We provide a
description of the algorithm, prove its convergence, and validate its
performance and solution tracking capabilities with numerical experiments.% As a byproduct, our convergence proof for the non-smooth DASF algorithm provides an alternative and leaner convergence proof for the original DASF algorithm, as a special case.

\end{abstract}


% Article structure
% I. Introduction
% - Distributed processing vs. fusion center
% - Related work: comparison of DASF with ADMM consensus, Block coordinate
%   descent and diffusion, NS smooth distributed solvers
% - Non-smooth improvement of DASF
% - Perform node selection : improvement on DMAXVAR, single problem to solve
% - Inexact fixed point solvers
% II. Problem statement
% - Distributed linear filtering problem
% - Time dependence of the problem
% III. Distributed Algorithm
% - Algorithm description
% - Generic case of aggregation with any subspace constraint
% - Convergence analysis
% IV. Inexact fixed point solvers
% - Algorithm description
% - Convergence analysis
% V. Node selection in multi-task networls
% VI. Numerical results
% VII. Conclusion
\section{Introduction}
\label{sec:introduction}

% Context:
% - Edge computing, stress on the network
% - Wireless sensor networks with limited bandwidth (in part due to power
%   constraints)
% - Real time requirements
% - Distributed processing vs. fusion center (which can be a datacenter)
% - Compress/Fuse then send. Compression via adaptive filter.

\IEEEPARstart{T}{he} advent of Cloud computing and the so-called Big Data era
has led to a significant increase in the amount of data that is being
generated, and subsequently processed.
The vision put forth by the Cloud is that data is generated at the ``edge'', and
compute is located at the ``center'' \cite{cao2020overview}
, making it the most extreme form of
centralized computing. In the context
of Wireless Sensor Networks (WSNs), the center is usually referred to as the fusion
center (FC) and is responsible for the analysis and processing of the signals
gathered by a possibly large set of wirelessly connected sensor nodes. Although conceptually simple, this ``data
centralization'' approach comes with several
challenges. First, it requires significant bandwidth to transfer data from the
sensing devices to the central computing location \cite{Bertrand2015}
. This can be
particularly problematic in the case of low-power WSNs, where centralized data aggregation can have a significant
impact on the
battery life of the individual sensor nodes, making long term or remote deployments impractical. Second, the back and
forth communication between the computing device and sensing devices can prevent
real-time computations to be performed due to increased network latency \cite{cao2020overview}. Time-critical tasks such as speech enhancement in acoustic sensor networks \cite{bertrand2010adaptive}, auditory attention decoding in EEG sensor networks \cite{narayanan2021eeg,narayanan2020optimal} or target detection in decentralized radar systems \cite{goodman2007optimum},
require
real-time data analysis to be of any use. Finally, the FC constitutes a single point of
failure \cite{Haykin2010}, which can lead to loss of service in case of malfunction. This
motivates the development of distributed algorithms that can process the data
both locally and 
in real-time, at the edge of the network, or collaboratively
compute optimally compressed representations of the data that can be efficiently
offloaded to an external computing device, at a much lower cost than the
transmission of the raw sensor data.

Distributed datasets as those sensed by sensor networks can be broadly classified into two categories: those
with distributed features, and those with distributed samples. The latter
is usually less challenging to process, as the objective function can typically be decomposed as a sum of local node-specific objective functions that only require the samples from one node. In this case, the distributed task can often be solved by first
processing the samples of each node locally and independently from the other
nodes, and then aggregating the intermediate local results (which typically consist of a parameter vector, rather than individual time samples). Typical examples of
algorithms suited for datasets with distributed samples are those put forth by federated learning
\cite{li2020federated}, the Alternating Direction Method of Multipliers (ADMM)
\cite{boyd2011distributed}, and consensus and diffusion strategies\cite{lopes2008diffusion,
sayed2014adaptation}, amongst others. In the case of distributed features, the objective function cannot be decomposed into a sum of local objectives at the individual nodes, making decentralized processing significantly more challenging. For example, in the context of spatial filtering in WSNs, the goal is to linearly combine the different sensor channels of all nodes in an optimal data-driven fashion, which requires measurements of the signal covariance across all node pairs in the WSN. \cite{liu2022fedbcd} describes an algorithm to process such datasets, but is limited to solving smooth and unconstrained optimization problems, in star-topology networks only, contrarily to this paper.

This work extends the Distributed Adaptive Signal Fusion (DASF) algorithm
\cite{musluoglu2022unified_p1, musluoglu2022unified_p2}, which was originally
designed to solve such distributed features problems, in particular in the context of spatial filtering in WSNs, where the sensor channels are distributed across the nodes of a WSN. Specifically, the DASF framework provides a generic ``meta-algorithm'' which computes adaptive spatial filters whose coefficients are the solutions of some smooth optimization problem, where the latter defines the optimal centralized spatial filter based on the network-wide signal statistics (which are assumed to be unknown at start-up). The DASF algorithm relies on the exchange of
linearly compressed views of the nodes' data, which are then used by
each node to locally solve a subproblem preserving the  original problem structure,
and iteratively producing a better estimate of the optimal filter coefficients. By spreading iterations of the algorithm over different sample batches, DASF is able to adaptively track the optimal filters based on their evolving statistics, which are estimated online using the most recently collected samples.
The DASF framework is applicable to a wide class of spatial filtering problems, including the trace ratio problem \cite{wang2007trace}, principal component analysis \cite{Hotelling1933}, generalized eigenvalue problems \cite{Golub2013}, minimum mean squared error filtering, minimum variance beamforming \cite{van1988beamforming}, and single and multi-view canonical correlation analysis \cite{Carroll1968}.
However, the DASF algorithm requires the optimization
objective to be smooth, which notably prevents the use of the $\ell_1$ norm,
which is often used in signal processing to encourage sparsity in the solutions. Our focus in
this work, is the extension of the DASF algorithm to non-smooth problems, with the side goal of performing $\ell_1$-induced node or sensor selection alongside a given filtering task (rather than performing an a priori node selection based on some task-agnostic criterion as in, e.g., \cite{hovine2022maxvar}).

Our contribution is three-fold. Firstly, we propose an extension of the original
DASF algorithm to certain classes of \emph{non-smooth} adaptive spatial filtering problems, referred to as non-smooth DASF (NS-DASF). Secondly, we provide a convergence and optimality proof for the NS-DASF algorithm based on milder assumptions than the original DASF algorithm. The new assumptions and optimality in the case of non-smooth problems lead to a very different proof strategy compared to the convergence proof of the smooth version of DASF.  %As a byproduct, this new convergence proof results in an alternative (and leaner) convergence proof for the original smooth DASF algorithm.
Finally, we
apply our algorithm to the problem of node selection in WSNs, where only a subset
of the nodes is required to contribute to the filtering task. 
In this paper, we show via numerical experiments that the problem of (distributed) node selection can be
solved concurrently with the filtering task, by the addition of an appropriate
regularizer to the optimization problem.

% In our previous work \cite{hovine2022maxvar}, we tackled the problem of node selection as a network partitioning problem,
%where the network was subdivided according to the correlation between the
%signals sensed by different subsets of nodes. This approach had the drawback of
%requiring an additional task to be solved prior to the actual filtering task of
%interest. 

%Third, we demonstrate with a numerical experiment that this extension equips the algorithm with the ability to perform node selection in conjunction with the search for the optimal network-wide spatial filter in a distributed setting.

The outline of the paper is as follows. In Section \ref{sec:problem} we formalize the scope of DASF and NS-DASF in a WSN context. In Section \ref{sec:distralg} we describe the NS-DASF algorithm. We first introduce the simpler case of fully-connected connected networks, before extending the description to arbitrary network topologies. In Section \ref{sec:convergence}, we prove the convergence and optimality of NS-DASF. Section \ref{sec:simulations} describes several numerical experiments, and Section \ref{sec:discussion} concludes the paper with a brief discussion.

\section{Problem Statement}
\label{sec:problem}

We consider a network of $K$ nodes with labels in $\mc{K}=\{1,\ldots,K\}$. Each node $k$ senses
an $M_k$-channel stochastic signal $\y_k(t)$ with values in
$\mathbb{R}^{M_k}$ for each sample $t\in\mathbb{Z}$. We
denote the network-wide $M$-channel signal with values in $\mathbb{R}^{M}$ as $\y(t) \triangleq
[\y_1(t)^T,\ldots,\y_K(t)^T]^T$, where $M\triangleq\sum_k M_k$.  
In this paper, we focus on the distributed
computation of an adaptive $M$-inputs $Q$-outputs spatial filter $\X \in \mathbb{R}^{M\times Q}$, that is optimal in some sense, and structured as  $\X \triangleq
[\X_1^T,\dots,\X_K^T]^T$, where $\X_k$ is the $k$-th block of $\X$, corresponding to the filter
associated with $\y_k$.

\subsection{Scope of the Original DASF Framework}
The original smooth version of DASF \cite{musluoglu2022unified_p1, musluoglu2022unified_p2}, applies to filtering problems of the form
\begin{equation}
    \label{eq:generic_problem_smooth}
    \begin{split}
    \X^\star & \in \argmin_{\X\in\mathbb{R}^{M\times Q}} \; \varphi(\X^T \bm{y}(t), \X^T\B)\\
    \st &\forall j\in\mc{J}_I,\; \eta_{j}(\X^T\y(t), \X^T \C_j ) \leq 0,\\
            &  \forall j\in\mc{J}_E, \;\eta_{j}(\X^T\y(t), \X^T \C_j ) = 0.
    \end{split}
\end{equation}
where the matrices
$\B$ and $\C_{j}$\footnote{The non sequitur notation is chosen to stay consistent with the notation introduced in \cite{musluoglu2022unified_p1}, where $C_k$ has another meaning, used later in this paper.} are
deterministic matrices known by every node, $\varphi$ is a
smooth real-valued function encoding some design objective for the filter
output, $\mc{J}_I$ and $\mc{J}_E$ are the sets of inequality and equality constraints indices, respectively, and $\eta_{j}$ are smooth
functions enforcing some hard
constraints on the filter outputs and/or filter coefficients. As $\bm{y}(t)$ is
a stochastic signal, we assume that the aforementioned functions implicitly
contain an operator extracting some statistics from the signal, such as e.g. an
expectation or covariance operator, hence keeping the problem deterministic\footnote{In order to be perfectly rigorous, we could have defined the domain of the above functions as a subset of some Hilbert space of random signals. This would however have introduced unnecessary complexity, as for all intents and purposes, the random signals could be replaced by sample matrices and leave our developments mostly unchanged.}.  In order for the nodes to be able to evaluate, or at least approximate those quantities, we assume that $\y(t)$ is ergodic, i.e. its statistics can be evaluated using sample averages. We furthermore assume that $\y(t)$ is short-time stationary and that its statistics change sufficiently slowly such that they can be tracked by the updates of the DASF algorithm. We emphasize this fact by dropping the time index $t$ for most of the remainder of this paper.
%\remove{LCMV filters \cite{van1988beamforming} are a well-known example of filter matching the structure of \eqref{eq:generic_problem_smooth}:
%\begin{align}
%        \X^\star \in &\argmin_\X \X^T\E{\y\y^T}\X\\
%               &\st \X^T\C = H.
%\end{align}}
\begin{align}
        \X^\star \in &\argmax_\X \X^T\E{\y\y^T}\X\\
               &\st \X^T\X = {I}
\end{align}
 is a typical instance of \eqref{eq:generic_problem_smooth}. It produces a filter that extracts the principal components of $\y$. In this particular case $\B=0$, and the sole constraint is
\begin{equation}
    \eta(\X^T\y, \X^T\C) = X^T\C X = I,
\end{equation}
with $\C=I$. Although the $\C$ matrix looks superfluous here, it plays an important algorithmic role, and allows DASF to solve \eqref{eq:generic_problem_smooth} by solving a sequence of lower-dimensional versions of \eqref{eq:generic_problem_smooth}, but where $\C$ (or $\B$ when it is non-zero) is not equal to the identity matrix anymore. %See \cite{musluoglu2022unified_p1} for the full details.

\subsection{Extended Scope of the NS-DASF Framework}

The NS-DASF algorithm deals with filters that are solutions of optimization problems of the form 

\begin{equation}
    \label{eq:generic_problem}
    \begin{split}
    \X^\star & \in \argmin_{\X\in\mathbb{R}^{M\times Q}} \; \varphi(\X^T \bm{y}(t), \X^T\B) + \gamma(\X^T\A)\\
        \st   \forall k\in \mc{K},\;&\forall j\in\mc{J}_I^k,\; \eta_{j}(\X_k^T\y_k(t), \X^T_k \C_{j,k} ) \leq 0,\\
            &  \forall j\in\mc{J}_E^k, \;\eta_{j}(\X_k^T\y_k(t), \X^T_k \C_{j,k} ) = 0
    \end{split}
\end{equation}
where the matrices $\A$
and $\C_{j,k}$\footnote{Most problems are usually formulated with $\gamma(A\X)$ rather than $\gamma(\X^TA)$. This notation was chosen to stay consistent with the $\X^T\cdot$ structure of the remaining functions involved in the problem.} are, similarly to $\B$,
 deterministic matrices known by every node, $\gamma$ is a possibly non-smooth but convex real-valued function encoding some
soft-constraints on the filter coefficients, and the $\eta_{j}$'s are smooth
functions now describing \emph{per-node}
constraints on the filter. Note that the
block-separability of the constraints is specific to the non-smooth version of
DASF, and is not required for its smooth counterpart. The $\eta_j$'s in \eqref{eq:generic_problem_smooth} can thus possibly introduce multiplicative coupling between the filter coefficients of different nodes, but the $\eta_j$'s in \eqref{eq:generic_problem} can only depend on a single block $\X_k$.

In order to ensure optimality (see Section
\ref{sec:convergence}), we also require
$\gamma$  to be \emph{per-node} block separable, i.e. there exist functions
$\gamma_k$ such that
\begin{equation}
\label{eq:per_node_block_separable}
\gamma(\X^T\A) = \sum_{k\in\mc{K}} \gamma_k(\X_k^T\A_k)
\end{equation}
This implies that $\A$ must be a block diagonal matrix, whose
blocks we denote $\A_k$. Typical examples of functions satisfying this
property are the weighted $\ell_1$ and $\ell_{2,1}$ norms, where the later is typically used
to introduce group-sparsity in an optimization model. Note that the smooth function $\varphi$ is not required to be block-separable. 

%In order to provide a unifying approach, our proof framework will also cover the optimality of the original smooth version of DASF \cite{musluoglu2022unified_p1, musluoglu2022unified_p2}, that applied to problems of the form

We denote the parametric optimization problem defined in \eqref{eq:generic_problem} as $\mathbb{P}(\y, \A,\B, \C)$, where $\mc{D}$ denotes the collection (i.e. set) of matrices $\C_{j,k}$.

%Figure \ref{fig:example_network} illustrates an example
%4-nodes network with associated filters and data. 

By imposing the proper structure on $\bm{y}(t)$, $\varphi$, $\gamma$,
and the constraints, a wide range of
problems can be cast to the form of \eqref{eq:generic_problem}. 
For example, the problem
\begin{equation}
    \label{eq:sumcorr}
        \begin{split}
                \max_{\X} &\; \tr{\X^T \E{\bm{y}(t)\bm{y}(t)^T}\X}\\
                \st &\; \X_k^T \E{\bm{y}_k(t)\bm{y}_k(t)^T}\X_k ={I}_Q\quad\forall k\in\mc{K}
        \end{split}
\end{equation}
where ${I}_Q$ is the $Q$-dimensional identity matrix, and $\tr{\cdot}$ and $\E{\cdot}$ denote the trace and expectation operators,
respectively, is typically referred to as the SUMCORR formulation of generalized
canonical correlation analysis (GCCA) \cite{Kettenring1971, Kanatsoulis2018, Soerensen2021} and can be cast in the
form \eqref{eq:generic_problem}. By adding $\gamma(\X)
= \sum_k\norm{\X_k}_{F}$  to the objective function of \eqref{eq:sumcorr}, where $\norm{\cdot}_{F}$ denotes the
Frobenius norm, we obtain a sparse version of SUMCORR, which encourages
a subset of the channels to be used. 
%This problem has been solved in a
%distributed fashion by \cite{Kanatsoulis2018}. Our proposed algorithmic
%framework will result in an alternative distributed algorithm, which we will
%compare with \cite{Kanatsoulis2018} in Section \ref{sec:simulations}. 
As another example, the following problem can be viewed as a sparse Wiener filtering problem \cite{vaseghi1996wiener} with additional power constraints on the per-node filter outputs:
\begin{equation}
    \label{eq:wiener}
        \begin{split}
                \min_{\X} &\; \E{\norm{\X^T\y(t)-\bm{d}(t)}_F^2} + \sum_{k}
                \norm{\X_k}_{F}\\
                \st &\; \E{\norm{\X_k^T
        \bm{y}_k(t)}_F^2} \leq P_k\quad\forall k\in\mc{K}
        \end{split}
\end{equation}
where $P_k$ are scalars denoting some power constraint on
the filter output, and
$\bm{d}(t)$ is a known target signal taking values in $\mathbb{R}^{Q}$. Note that these are only two examples of the many problems
that fit in the proposed non-smooth DASF framework. 

Our goal is to efficiently track a solution
$\X^\star \in \mathbb{R}^{M\times Q}$ of
\eqref{eq:generic_problem} and the corresponding filter output $\X^{\star
T}\bm{y}(t)$. 
Although the optimal filter $\X^\star$ is required, we are typically more interested in the output signals of the spatial filter, i.e., in the $Q$-dimensional filtered output $\z(t) \triangleq \X^{\star T}\y(t)$ for each sample time $t$, which could, for example, correspond to a denoised speech signal that must be available to the network at any time. The DASF algorithm must therefore be such that both the optimal filter and the filtered signal $\z(t)$ can be computed in a bandwidth efficient manner.

The NS-DASF algorithm assumes that a centralized solver that would be able to find the solution of \eqref{eq:generic_problem} if all data would be known at a fusion center is available. Similarly to the original DASF algorithm, the NS-DASF algorithm will use this solver to compute the solution of lower-dimensional versions of \eqref{eq:generic_problem} that are available at individual nodes. We also
assume that computing a solution of this local lower dimensional problem is cheap in
comparison to the cost of sharing the data $\bm{y}(t)$, which motivates the
design of a distributed algorithm that can compute $\X^\star$ and $\bm{z}(t)$ while also limiting
the amount of data that needs to be exchanged between the nodes, by relying on
local computations instead. This is a reasonable assumption, as it is well known that the wireless data exchange is typically an energy bottleneck in WSNs \cite{strypsteen2022bandwidth,Bertrand2015}.

\section{The Non-Smooth DASF Algorithm}
\label{sec:distralg}

In this section, we describe the non-smooth DASF (NS-DASF) algorithm in earnest,
i.e. we describe an iterative procedure to solve
\eqref{eq:generic_problem} in a distributed fashion, while also tracking the filtered output $\z(t)$ at each node. The procedure relies on
each node sending a compressed view of its local data to a given node, which we
call the \emph{updating node}, and whose role is assumed by a different node at each iteration. Based on the compressed data it received, the updating node will update the current estimate
of $\X^\star$. For the sake of an easier exposition, we first
introduce the algorithm in fully-connected networks before extending the description
to arbitrary network topologies at the end of the section.

\subsection{NS-DASF in Fully-Connected Networks}
\label{sec:algorithm_description}
%\remove{
%\input{fc_procedure.tex}
%}
The state of the algorithm at any iteration $i$ is characterized by the (initially random) current
estimate of the optimal solution $\X^i$ and the updating node index $q^i$ (we use the node index $q$ without any iteration index to refer to the updating node when the iteration is clear from the context).
The algorithmic procedure is essentially the same as the one described in \cite{musluoglu2022unified_p1} for its smooth counterpart, except for the
handling of the term $\gamma$ and the resulting local problems solved by each
updating node. However, the convergence analysis (and in particular the optimality results) is substantially impacted by the addition of this
non-smooth term (see Section
\ref{sec:convergence}).

Each iteration is divided into three phases:
\paragraph*{(i) Data Aggregation} Each node $k$ collects $N$ new samples of $\y_k$ and sends a block of $N$ $Q$-dimensional
\emph{compressed} samples of
\begin{equation}
    \label{eq:compressed_data}
            \cy_k^i \triangleq \X_k^{iT}\y_k
    \end{equation}
along with
    \begin{align}
    \label{eq:compressed_data_2}
            \cC^i_{j,k} &\triangleq \X_k^{iT}\C_{j,k}\\
            \cA_k^i &\triangleq X_k^{iT}\A_k    \\
            \cB_k^{iT} &\triangleq \X_k^{iT}\B_{k}
\end{align}
to the updating node $q$, where $\B_k$ is the block-row of $\B$ associated with $\X_k$. Note that $\X_k^i$ acts both as the compression matrix and as
the current estimate of the optimal filter. Upon reception of the compressed
data, the updating node $q$ constructs  the \emph{local} data
\begin{equation}
    \label{eq:local_data}
    \begin{split}
        \ly^i &= \begin{bmatrix}
            \y_q^T & \cy_1^{iT} & \cdots & \cy_{q-1}^{iT} & \cy_{q+1}^{iT} & \cdots &
            \cy_K^{iT}
        \end{bmatrix}^T,\\
            \lA^i &= \blkd(\A_q, \cA_1^i, \ldots,
    \cA_{q-1}^i, \cA_{q+1}^i, \ldots, \cA_K^i),
\textrm{ and }\\
           \lB^i &= \begin{bmatrix}
            \B_q^T & \cB_1^{iT} & \cdots & \cB_{q-1}^{iT} & \cB_{q+1}^{iT} & \cdots &
            \cB_K^{iT}
        \end{bmatrix}^T
        \end{split}
\end{equation}
adding node $q$'s own data to the aggregated data. Similarly to $\mc{D}$, we denote the collection of the $\cC_{j,k}^i$'s and the $\C_{j,q}$ as $\widetilde{\mc{D}}^i$.
One can already notice that the ``local'' data $\ly^i$, $\lA^i$, $\lB^i$ and $\widetilde{\mc{D}}^i$ live in a
subspace of the original data $\y$, $\A$, $\B$ and $\mc{D}$, and can be interpreted as a low-dimensional
``view'' (i.e. linear combination of the channels/rows) of the original data, where the view of the node's
own data is unaltered (i.e. uncompressed).
\paragraph*{(ii) Local Solution} In order to update the current estimate of the
optimal filter $\X^i$, the updating node $q$ computes a solution of the original
problem \eqref{eq:generic_problem}, but using the aggregated  data \eqref{eq:local_data} received in the previous step instead
of the original, global, data $\y$, $\A$, $\B$, and $\mc{D}$. More specifically, it solves $\mathbb{P}(\ly^i, \lA^i,\lB^i, \widetilde{\mc{D}}^i)$:
\begin{equation}
        \label{eq:local_problem}
    \begin{split}
        \lX^{\star} & \in \argmin_{\lX} \; \varphi(\lX^T \ly^i,\lX^T \lB^i) + \gamma(\lX^T\lA^i)\\
        \st  \forall   k\in \mc{K},\; &\forall j\in\mc{J}_I^k,\; \eta_{j}(\lX_k^T\cy_k^i, \lX_k^T\cC_{j,k}^i) \leq 0,\\
            &  \forall j\in\mc{J}_E^k, \;\eta_{j}(\lX_k^T\cy_k^i, \lX_k^T\cC_{j,k}^i) = 0
    \end{split}
\end{equation}
where we denote $\cy^i_q \triangleq \y_q$, $\cB^i_q \triangleq \B_q$ and $\cC^i_q \triangleq \C_q$ for the special case of the updating
node, and where $\lX^{\star}$ is partitioned as
\begin{equation}
\label{eq:block_struct_local_filter}
    \lX^{\star} = [\lX_q^{\star T}, \lX_1^{\star T}, \ldots, \lX_{q-1}^{\star T}, \lX_{q+1}^{\star T}, \ldots, \lX_K^{\star T}]^T.
\end{equation}
with each block associated with the corresponding blocks of the local data \eqref{eq:local_data}. Note that $\lX_q$ is an $M_q \times Q$ matrix, whereas all the other $\lX_k$'s are $Q \times Q$ matrices. Also note that because the node receives blocks of samples, it will not solve $\mathbb{P}(\ly^i, \lA^i,\lB^i, \widetilde{\mc{D}}^i)$ exactly, but an approximation thereof, where the implicit statistics involved in $\varphi$ and the $\eta_j$'s are approximated using sample averages across a batch of $N$ samples.

As \eqref{eq:local_problem} shares the structure of the original problem
\eqref{eq:generic_problem}, it can be solved using the same solver. It should be
noted that the dimension of this local problem is much smaller than the
original problem, and therefore cheaper to solve, making it amenable to run on devices with limited computing capabilities.% Specifically, the dimension of the global optimization variable $\X$ is $Q\times M$, while the dimension of the local variable $\lX$ at node $q$ is $Q\times (M_q + (K-1)Q)$.

\paragraph*{(iii) Parameters Update} 

Following the partitioning defined in \eqref{eq:block_struct_local_filter}, the updating node $q$ updates its own filter according to
\begin{equation}
\label{eq:update_rule_q}
\X_q^{i+1}\leftarrow \lX_q^{\star}
\end{equation}
and sends the appropriate blocks of
$\lX^{\star}$ to the other nodes, such that they can update their local filters according to
\begin{equation}
        \label{eq:update_rule_k}
\X_k^{i+1}\leftarrow \X_k^i\lX_k^{\star}.
\end{equation}
The above rule is related to \eqref{eq:compressed_data}-\eqref{eq:compressed_data_2}: At each node, the  signal $\y_k$ of a node can only be manipulated by the updating node ``through'' the current estimate of the filter $\X^i_k$, via the parametrization introduced by $\lX_k^{\star}$.% A schematic representation of this back and forth communication procedure is shown in Figure
%\ref{fig:fc_procedure}.

Upon completion of the above steps, the updating node role is
passed-on to another node\footnote{As will be discussed in Section \ref{sec:convergence}, the order does not matter as long as each node acts as an updating node an infinite number of times.} and another iteration begins,
using another batch of $N$ samples. At the end of each iteration, the updating node has access to an estimate of the output filtered signal  for the latest $N$-samples block:
\begin{equation}
        \z \approx \lX^{\star T}\ly^i = \X^{i+1T}\y.
\end{equation}
As a different batch of $N$ samples is used at each iteration, the (NS-)DASF algorithm produces an estimate of the filtered signal for each $N$-samples block, while at the same time improving the estimate of the optimal spatial filter $\X^\star$, such that each new block of the filtered signal is closer to the desired filtered signal (under the stationarity assumption). In other words, (NS-)DASF acts as a time-recursive block-adaptive filter, which continually adapts itself to the (possibly changing) statistics of $\y(t)$ \cite{musluoglu2022unified_p1}.

The full algorithm description is given by Algorithm \ref{alg:fc}.% Note that the algorithm is almost identical to the original DASF algorithm in \cite{musluoglu2022unified_p1} except for the additional transmissions of the $\lA_k^i$.

\SetKwFor{Node}{At node}{}{}
\SetKwFor{Loop}{loop}{}{}
\begin{algorithm}[t]
        \footnotesize
        \Begin{
                $i\gets 0$, $q\gets 1$, Randomly initialize $\X^0$\\
                \Loop{}{
                        \For{$k\in\mc{K}\smallsetminus\{q\}$}{
                                \Node{$k$} {
                                        Collect a new batch of $N$ samples of
                                        $\y_k(t)$ and send the compressed
                                        samples
                                        $\cy_k^i(t)=\X_k^{iT}\y_k(t)$
                                        along with $\Al^i_k=\X^{iT}_kA_k$, $\cB^i_k=\X^{iT}_k\cB_k$ and
                                        $\cC_k^i = \X^{iT} \C_k$ to node $q$.
                                }
                        }
                        \Node{$q$}{
                                Obtain $\lX^{\star}$ by solving and selecting any solution of
                                $\mathbb{P}(\ly^i(t),\lA^i,\cB^i_k, \widetilde{\mc{D}}^i)$ (see \eqref{eq:local_problem}).\\                                Extract the $\lX_k^{\star}$'s
                                from ${\lX}^{\star}$ according to \eqref{eq:block_struct_local_filter}.\\
                                $\X_q^{i+1}\gets \lX_q^{\star}$\\
                                \For{$k\in\mathcal{K}\smallsetminus\{q\}$}{

                                        Send $\lX_k^{\star}$ to node $k$.\\
                                        \Node{$k$}{
                                                $\X^{i+1}_k \gets
                                                \X^{i}_k\lX_k^{\star}$\\
                                        }
                                }

                        }
                        $i\gets i+1$, $q\gets (q + 1) \mod K$

                }
        }
        \caption{NS-DASF algorithm in fully-connected networks.}
        \label{alg:fc}
\end{algorithm}

\subsection{NS-DASF in Arbitrary Network Topologies}
\label{sec:arbitrary}

A fully-connected network topology allows the updating node to receive the
compressed data of every other node directly. In an arbitrary network topology,
the updating node can only receive data from its neighbors. Applying the same
procedure as in the fully-connected case, i.e. requiring the nodes to relay the compressed data of their
neighbors to the updating node, would result in a significant communication
overhead, most extreme in the case of line topologies. To avoid this, it was proposed in \cite{musluoglu2022unified_p1} to construct at each iteration a spanning tree that is rooted at the updating node, and let each node fuse (i.e. sum) the compressed data of its neighbors before relaying it to the neighbor closest to the
updating node. In other words, the updating
node will receive some linear combination of the compressed data of the rest of the network.
 With this interpretation in mind, the procedure described for fully
connected networks can readily be applied, considering each branch at a given iteration \emph{as if
        it were a single
node}. The local variables $\X_k$ of each node in a branch are therefore updated with the
same linear transformation (i.e. there is a single matrix $\lX_{(\cdot)}$ per
branch).

As we have required the constraints to be block separable, we would in practice still need
to forward all the compressed data $\X^{Ti}_k \y_k$ in order for the updating
node to be able to evaluate each $\eta_j(\X^{Ti}_k \y_k, \X^{Ti}\C_k )$. As this would require
expensive data relaying, we require the constraints to depend only on the first or second order statistics of $\X^T_k\y_k$ in the arbitrary topology case. This allows the nodes to relay the compressed data $\X^{Ti}_k\C_k$ along with $\X^{Ti}_k\E{ \y_k\y^T_k}\X_k$ or $\X^{Ti}_k\E{ \y_k}$ instead of a batch of $N$ samples of $\X^{Ti}_k \y_k$, which is of much larger dimension than the compressed statistics
(which have a dimension of the same order of magnitude as the parameter update matrices $\lX$). For example, to handle  constraints of the form $$\X_k^T\E{\y_k  \y^T_k}\X_k=I_Q,$$ the nodes
would relay (sample-averaged estimates of) $\cR_k^i \triangleq\X^{iT}_k\E{\y_k \y^T_k}\X^i_k$ itself and still be able to evaluate the
constraints, as then the local constraints in \eqref{eq:local_problem} would be of the form
\begin{equation}
\lX^T_k\cR_k^i \lX_k=I_Q.
\end{equation}
%Note that the constraints depend on $\cy_k$, rather than $\ly_k$.

\begin{figure}[t]
    \scalebox{0.8}{
        \tikzstyle{node}=[circle,draw=none,fill=blue!20,minimum size=1.5em,inner sep=0pt]
        \tikzstyle{updating}=[circle,draw=none,fill=red!40,minimum size=1.5em,inner sep=0pt]
    \tikzstyle{halo}=[circle, minimum size=1cm]
    \centering
 %   \subfloat[Non fully-connected network]{
 %       \begin{tikzpicture}
 %           % A network of nodes
 %           \node[node] (k1) at (0,0) {$1$};
 %           \node[node] (k2) at (1.5,-1.5) {$2$};
 %           \node[node] (k3) at (0,-3) {$3$};
 %           \node[node] (k4) at (-1.5,-1.5) {$4$};

 %           \draw[dashed] (k1) -- (k2);
 %           \draw[dashed] (k1) -- (k4);
 %           \draw[dashed] (k3) -- (k2);
 %           \draw[dashed] (k3) -- (k4);

 %           % same structure to the left
 %           \node[node] (k5) at (-3,0) {$5$};
 %           \node[node] (k7) at (-3,-3) {$7$};
 %           \node[node] (k6) at (-4.5,-1.5) {$6$};

 %           \draw[dashed] (k4) -- (k5);
 %           \draw[dashed] (k4) -- (k7);
 %           \draw[dashed] (k6) -- (k5);
 %           \draw[dashed] (k6) -- (k7);
 %       \end{tikzpicture}
 %   }
 %   \hfill
 %   \subfloat[Aggregation in a spanning tree]{
        \begin{tikzpicture}
            % A network of nodes
            \node[updating] (k1) at (0,0) {$1$};
            \node[node] (k2) at (1.5,-1.5) {$2$};
            \node[node] (k3) at (0,-3) {$3$};
            \node[node] (k4) at (-1.5,-1.5) {$4$};

            % same structure to the left
            \node[node] (k5) at (-3,0) {$5$};
            \node[node] (k7) at (-3,-3) {$7$};
            \node[node] (k6) at (-4.5,-1.5) {$6$};

            % Aggregation to k1
            \draw[-Latex] (k2) to node[midway, above, sloped] (agg2) {$\cy_{21}$} (k1);
            \draw[-Latex] (k4) to node[midway, above, sloped] (agg4) {$\cy_{41}$} (k1);
            \draw[-Latex] (k3) to node[midway, above, sloped] {$\X^{iT}_3\y_3$} (k2);
            \draw[-Latex] (k7)  to node[midway, above, sloped] {$\X^{iT}_7\y_7$} (k4);
            \draw[-Latex] (k6) to node[midway, above, sloped] {$\X^{iT}_6\y_6$} (k5);
            \draw[-Latex] (k5) to node[midway, above, sloped] (agg5) {$\cy_{54}$} (k4);

            \node (label_agg) at ($(agg5)!0.5!(agg2) + (0,2)$) {Fused  $Q$-dimensional data};
            \draw[->] (label_agg) to[bend left] (agg2);
            \draw[->] (label_agg) to[bend right] (agg4);
            \draw[->] (label_agg) to[bend right] (agg5);

        \begin{pgfonlayer}{background}
            \draw[fill=green!10, draw=green] ($(k6.west)+(-0.2,0)$) to[closed, curve through={($(k5.north)+(0,0.2)$) .. ($(k4.east)+(0.2,0)$) .. ($(k7.south)+(0,-0.2)$) } ] ($(k6.south)+(0,-0.2)$);
            \draw[fill=orange!10, draw=orange] ($(k3.west)+(-0.2,0)$) to[closed, curve through={($(k2.west)+(-0.5,0)$) .. ($(k2.north)+(0,0.2)$) .. ($(k2.east)+(0.2,0)$) ($(k3.east)+(0.2,0)$) } ]  ($(k3.south)+(0,-0.2)$);
            \node at ($(k7)+(-1.75,0.5)$) {$\mathcal{B}_{41}$};
            \node at ($0.5*(k3)+0.5*(k2)+(1.5,0)$) {$\mathcal{B}_{21}$};*
        \end{pgfonlayer}

        \end{tikzpicture}
    }
   % }
    \caption{Example of an aggregation scheme in a spanning tree rooted at node $1$. Transmission of $\cR_k^i$, $\cB_k^i$ $\cC_k^i$ and $\cA_k^i$ omitted.}
    \label{fig:agg_tree}
\end{figure}
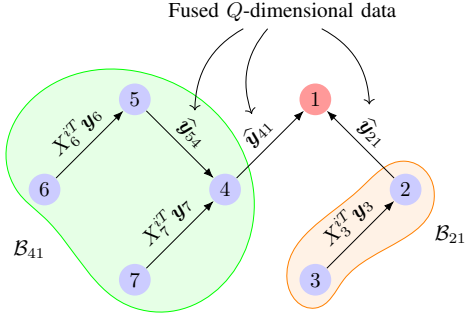

In what follows, we denote the per-node covariance matrix $\E{\y_k  \y_k^T}$ as $R_k$, and the per-node first order statistics $\E{\y_k}$ as $r_k$. % We emphasize the dependence of the constraints on $R_k$ by explicitly denoting the constraint set $\mc{X}'(R,\C)$, where 
%and Similarly to $\C$, we denote as $R$ the concatenation of the individual $R_k$. 
The procedure in arbitrary topologies is as follows.
\paragraph*{(i) Data aggregation}
A spanning tree rooted at the updating node and preserving the links with its neighbors is computed in a distributed fashion, using e.g.
\cite[Algorithm 4]{hovine2022maxvar}. We denote the set of nodes in the subtree (i.e. branch) containing $k$ and obtained by removing the link between $k$ and $q$ as $\mc{B}_{kq}$ (see Figure \ref{fig:agg_tree} for an example). The compressed data  $\cC^i$ and
 $\Al_k$ are forwarded to the updating node. 
Similarly, each node $k$ computes and forwards its compressed first and second-order statistics 
\begin{gather}
        \cR_k^i = \X^{Ti}_k R_k\X_k^i,\\
        \ccr_k^i \triangleq \X^{Ti}_k r_k
\end{gather}
based on the latest available batch of samples. We denote the compressed data associated with subtree $\mc{B}_{kq}$ as 
 \begin{equation}
         \label{eq:compressed_data_arbitrary}
         \begin{aligned}
         \cy^i_{kq}=\sum_{l\in\mc{B}_{kq}} \cy^i_l, \textrm{ and}\\
         \cB^i_{kq}=\sum_{l\in\mc{B}_{kq}} \cB^i_l.
 \end{aligned}
\end{equation}
This can be computed recursively by having each node $k$ in the subtree sum the data it receives from its children, and then forward the result to its parent.
This ensures that the dimension of the data sent by each node stays equal to $Q$, independently of the network size and topology, making the communication of the compressed $N$-sample batches fully scalable.
The full aggregation procedure is formally described by Algorithm \ref{alg:data_aggregation} and an illustrative example is shown in Figure \ref{fig:agg_tree}. Upon completion,  the updating
node $q$ has access to the compressed data $\cy^i_{kq}$ of each of its neighbors $n\in\mc{N}_q$, and the set of compressed matrices  $\cR^i_k$, $\ccr_k^i$, $\cC^i_{j,k}$ and
 $\Al_k$ for every node in the network.

 The updating node constructs the local data $\ly^i$ as
\begin{equation}
    \label{eq:local_data_arb}
    \begin{split}
            \ly^i &\triangleq \begin{bmatrix}
                \y_q^T & \cy_{n_1k}^{iT} & \cdots & \cy_{n_Lk}^{iT}
        \end{bmatrix}^T ,     \\ 
           \lB^i &\triangleq \begin{bmatrix}
            \B_q^T & \cB_{n_1k}^{iT} & \cdots  & \cB_{n_Lk}^{iT}
        \end{bmatrix}^T
\end{split}
\end{equation}
        where $\{n_1,\dots,n_L\} = \mc{N}_q$ are the neighbors of node $q$.
        The matrix $\lA^i$, is constructed as in the fully connected case \eqref{eq:local_data}.
\paragraph*{(ii) Local solution}
The updating node solves
\begin{equation}
        \label{eq:local_problem_arbitrary}
        \begin{split}
                \lX^\star\in
                \argmin_{\lX} &\;\varphi(\lX^T\ly^i, \lX^T \lB^i ) + \gamma(\lA^i\lX)\\
                \st &  \forall  \; n\in \mc{N}_q, \forall l\in\mc{B}_{nq}\\
            &\forall j\in\mc{J}_I^l,\; \eta_{j}(\lX_n^T\cy_l^i, \lX^T_n \cC_{j,l}^i ) \leq 0,\\
            &  \forall j\in\mc{J}_E^l, \;\eta_{j}(\lX_n^T\cy_l^i, \lX^T_n \cC_{j,l}^i ) = 0,
                             % &\st \lX\in \mc{X}(\lR^i,\lC^i)
        \end{split}
\end{equation}
where the dependence on the $\cR_k^i$ and $\ccr_k^i$ is implicitly embedded in the $\eta_j$.

\paragraph*{(iii) Parameters Update}
$\lX^{\star}$ is now partitioned
correspondingly to \eqref{eq:local_data_arb} as
\begin{equation}
        \label{eq:partitioned_lX_arbitrary}
        \lX^{\star} \triangleq \begin{bmatrix}
                \X_q^{\star T},
                \lX_{n_1}^{\star T},
                \dots,
                \lX_{n_L}^{\star T}
        \end{bmatrix}^T.
\end{equation}
Similarly to the fully-connected case, the updating node updates its filter
with $\X^{\star}_q$ and for each branch $\mc{B}_{n_lq}$, the nodes all update their filters according to
$\X^{i+1}_k\leftarrow \X_k^{i}\lX^{\star}_{n_l}$ for all $k\in\mc{B}_{n_lq}$ (Note that $\lX^{\star}_{n_l}$ is the same for every node in the branch).  The full algorithmic procedure
is described by Algorithm \ref{alg:arb}.

\begin{remark}
        The dimension of the local problem \eqref{eq:local_problem_arbitrary} directly depends on the number of neighbors of the updating node kept when building the spanning tree. Indeed, the local optimization variable will have dimension $(Q|\mc{N}_q|+M_q)\times Q$. We could in practice build a spanning tree ignoring some of the links between the updating node and its neighbors, but this would not yield any savings in required bandwidth (every node still needs to forward its data to some node), and it would be at the expense of convergence speed, as the available degrees of freedom available when minimizing the local problems \eqref{eq:local_problem_arbitrary} would then be lower.
\end{remark}

\SetKwFor{Node}{At node}{}{}
\SetKwFor{Loop}{loop}{}{}
\SetKwInOut{Input}{input}
\begin{algorithm}[t]
 \footnotesize
        \Input{Parent node $p_k$ and set of childrens $\mathcal{C}_k$ for each node $k\neq q$ (the parent node is the neighbor closest to the updating node $q$, $\mc{C}_k=\emptyset$ for leaf nodes)}
        \Begin{
                \Node{$k$}  {
                        Collect a new batch of samples of $\y_k(t)$\\
                        Wait for the aggregate compressed signals received from children $\cy_{lk}^i$ along with the sets of compressed matrices $\{\cC_{j,m}^i\}_{m\in\mc{B}_{lk}}$,  $\{\cA_m^i\}_{m\in\mc{B}_{lk}}$, $\{\cR_m^i\}_{m\in\mc{B}_{lk}}$ and $\{\ccr_m^i\}_{m\in\mc{B}_{lk}}$ for
                        $l\in\mc{C}_k$ \\
                                               Send $\cy^i_{kp_k}=\cy^i_k + \sum_{l\in\mc{C}_k}
                                               \cy^i_{lk}$ to $p_k$ and similarly for $\cB^i_{kp_k}$ \\
                        Send $\{\cC_{j,k}^i\}\cup_{l}\{\cC_{j,m}^i\}_{m\in\mc{B}_{lk}}$, $\{\cA_k^i\}\cup_{l}\{\cA_m^i\}_{m\in\mc{B}_{lk}}$, $\{\cR_k^i\}\cup_{l}\{\cR_m^i\}_{m\in\mc{B}_{lk}}$ and  $\{\ccr_k^i\}\cup_{l}\{\ccr_m^i\}_{m\in\mc{B}_{lk}}$ to $p_k$ \\

                }
        }
        \caption{Recursive aggregation procedure in a tree-topology network rooted at node $q$.}
        \label{alg:data_aggregation}
\end{algorithm}

\SetKwFor{Node}{At node}{}{}
\SetKwFor{Loop}{loop}{}{}
\begin{algorithm}[t]
        \footnotesize
        \Begin{
                $i\gets 0$, $q\gets 1$, Randomly initialize $\bm{X}^0$\\
                \Loop{}{
                        Construct a spanning tree rooted at node $q$.\\
                        Aggregate the data according to Alg.
                        \ref{alg:data_aggregation}.\\
                        \Node{$q$}{
                                Obtain $\lX^\star$ by solving and selecting any solution of
                                \eqref{eq:local_problem_arbitrary}.\\
                                Extract the $\lX_k^{\star}$'s
                                from ${\lX}^{\star}$ according to
                                \eqref{eq:partitioned_lX_arbitrary}.\\
                                Update the filter of node $q$ with $\lX_q^{\star}$.\\
                                \For{$n\in\mathcal{N}_q$}{
                                        Send $\lX_{n}^{\star}$ to node $n$.\\
                                        \For{$l\in\mc{B}_{nq}$}{
                                                \Node{$l$}{
                                                        Wait for $\lX_{n}^{\star}$ and forward it to its children.\\
                                                        $\X_{l}^{i+1}\gets\X_l^{i}\lX_{n}^{\star}$\\
                                                }
                                        }
                                }

                        }
                        $i\gets i+1$, $q\gets (q + 1) \mod K$

                }
        }
        \caption{NS-DASF algorithm in arbitrary networks}
        \label{alg:arb}
\end{algorithm}

\section{Convergence and Optimality}
\label{sec:convergence}

\newcommand{\seq}[2]{(#1^i)_{i\in #2}}
\newcommand{\N}{\mathbb{N}}
\newcommand{\I}{\mc{I}}

In this section, we provide convergence and optimality results for the proposed
NS-DASF algorithm. The addition of the non-smooth term and the different technical assumptions under which convergence and optimality are obtained do not
allow for a straightforward extension of the convergence proofs of the original
DASF algorithm (see \cite{musluoglu2022unified_p2}). These proofs are the main contribution of this paper. In order to keep them accessible, the convergence and optimality analyses are first derived for fully-connected networks in Sections \ref{sec:rel_loc_glob} to \ref{sec:optimality_fixed}, while Section \ref{sec:ext_Cconv_arb} briefly describes how the results obtained for fully-connected networks can be extended to arbitrary network topologies.

Similarly to the original proofs described in \cite{musluoglu2022unified_p2} and as described earlier in Section \ref{sec:problem}, we must make two simplifying assumptions on $\bm{y}(t)$ to ensure that the optimal solution $\X^\star$ is time-independent and does not vary across iterations, which would make the convergence analysis of the algorithm mathematically intractable.

\paragraph*{Short-Time Stationarity}
We assume that the stochastic signal $\bm{y}(t)$ is stationary, as is typically assumed in the convergence analysis of adaptive filters. However, this should not be viewed as a practical limitation. In order to track the optimal solution $\X^\star$, it is in practice sufficient to assume that the changes in the statistics of $\y(t)$ are sufficiently slow compared to the convergence time of the algorithm. 

\paragraph*{Perfect  Estimation of the Signal Statistics}
As the functions involved in \eqref{eq:generic_problem} are real-valued, they
implicitly depend on the statistics of $\bm{y}(t)$. In practice, the actual
distribution of $\bm{y}(t)$ is unknown, and its statistics would typically be estimated from sample averages, i.e., the expected value operator in the examples \eqref{eq:sumcorr} and \eqref{eq:wiener} would be replaced with a sample average over a finite batch of samples. In the (NS-)DASF algorithm, these statistics are estimated on the most recent batch of $N$ samples that were transmitted by the nodes. However, the convergence proof assumes for mathematical tractability that these statistics are estimated perfectly, which means that the convergence results should be viewed as asymptotic results. We refer the
reader to the stochastic optimization literature for details on this topic (see, for example,
\cite{kim2015guide}).

\subsection{Relationship between Local and Global Problems (in Fully-Connected Networks)}
\label{sec:rel_loc_glob}

Before delving into the actual proof, we give some intuition about the relationship between global and local problems in the case of fully-connected networks (see Section \ref{sec:ext_Cconv_arb} for the extention to arbitrary network topologies). 

From \eqref{eq:compressed_data}-\eqref{eq:local_data}, it can be observed that $\ly^i$, $\lA^i$, $\lB^i$ and $\cC^i_{j,k}$ are linear (compressive) transformations of $\y$, $\A$, $\B$ and $\C_{j,k}$. Therefore, these variables can be related via a matrix transformation with some matrix $C^i_q$, such that $\ly^i=C^{iT}_q \y$ (and similarly for the other matrices, with the exception of $\C_{j,k}$, as explained below).  From \eqref{eq:compressed_data}-\eqref{eq:local_data}, it is clear that the matrix $C_q^i$ is constructed from the entries in $\X^i$, such that we can define $C_q^i$ as the result of a matrix-valued function $C_q(\cdot)$
such that $C_q^i\triangleq C_q(\X^i)$. With this notation  \eqref{eq:compressed_data}-\eqref{eq:local_data} can be written as\footnote{We purposedly omit the matrices $\C_{j,k}$ as their case is a bit different. Defining $\C_{j,k}'\triangleq[0\;\cdots\; \C_{j,k}^T\;\cdots\;0]^T$, the same relationship can be established by noting that the $k$-th block of $C_q(\X^i)D'_{j,k}$ is $\cC^i_{j,k}$ and $q$-th block  $C_q(\X^i)D'_{j,q}$ is $\C^i_{j,q}$,%$\X^T_kD_{j,k}=\X^TD_{j,k}'$ 
and extending the same reasoning used for $\y$. See the explicit structure of $C_q$ in the supplementary materials.}
\begin{equation}
        \begin{split}
        \label{eq:local_parametrization}
        \ly^i &= {C}_q(\X^i)^T\y,\quad  \\
        \lB^i &= {C}_q(\X^i)^T\B, \textrm{and}\\
        \lA^{i} &= {C}_q(\X^i)^T\A
\end{split}
\end{equation}
which is simply the expression in matrix form of the combination of \eqref{eq:compressed_data}-\eqref{eq:local_data}. Note that $C_q(\X)$ is here implicitly defined\footnote{The explicit description and structure of $C_q(X)$ is explained in the supplementary material for the interested reader.}. %Let us briefly look at the structure of $C_q(\X)$. 
%Therefore, the
%map $\X\mapsto {C}_q(\X)$ is trivially continuous.

An update rule naturally emerges from the parametrizations introduced by \eqref{eq:local_parametrization}, as by simple associativity,
\begin{equation}
        \label{eq:equiv_update_1}
        \begin{split}
        \lX^{\star T}\ly^i &= \lX^{\star T}\left(C_q(\X^i)^T\y\right)\\
                        &=\left(C_q(\X^i)\lX^{\star} \right)^T\y\\
                        &= \X^{i+1T}\y
\end{split}
\end{equation}
(and similarly for $\B$ and $\A$). The last equality follows from the structure of $C_q$, which flows directly from the update rules \eqref{eq:update_rule_q}-\eqref{eq:update_rule_k} (see supplementary materials). This means that we can also re-express \eqref{eq:update_rule_q}-\eqref{eq:update_rule_k} in matrix form as 
\begin{equation}
        \label{eq:equiv_update_2}
        \X^{i+1} \leftarrow C_q(\X^i)\lX^{\star}.
\end{equation}

%Armed with this notation, we can summarize the full algorithmic procedure by a
%single map 
%\begin{equation}
%    F_q(\X) \triangleq \argmin_{\X\in \mc{X}} \; L(\X)
%\end{equation}

This shows that the local problem \eqref{eq:local_problem} at node $q$ can be interpreted as a parametrized version
of the centralized problem \eqref{eq:generic_problem}, where the optimization
variable is now constrained to a smaller linear subspace defined by the column space of $C_q(\X^i)$. We define 
\begin{equation}
L(\X) \triangleq \varphi(\X^T\y, \X^T\B) + \gamma(\A\X)
\end{equation}
and the set  $\mc{X}$ as the set of feasible points of \eqref{eq:generic_problem}, allowing us to express the original global problem \eqref{eq:generic_problem} as
\begin{equation}
        \min_{\X\in\mathbb{R}^{M\times Q}} \; L(\X) \quad \st \; \X\in \mc{X}.
\end{equation}
Then, based on \eqref{eq:equiv_update_1}-\eqref{eq:equiv_update_2}, we find that the variable $\X^{i+1}$ that is obtained by solving \eqref{eq:local_problem} followed by the updates \eqref{eq:update_rule_q}-\eqref{eq:update_rule_k} must be a solution of the following problem:
\begin{equation}
    \label{eq:local_problem_param}
    \begin{split}
            \X^{i+1}\in&\argmin_{\X} \; L(\X)\\
        \st & \X \in \mc{X}\\
            & \X \in \range C_q(\X^i),
    \end{split}
\end{equation}
with $\range$ denoting the range or column space operator\footnote{With a slight abuse of notation, as $\X$ has $Q$ columns. We thus actually mean the $Q$-th Cartesian power of the column space.}. The block structure\footnote{Owing to the fact that $\X^{i+1}_k\leftarrow\X^i_k\lX^\star_k$ for every $k\neq q$, (see supplementary materials for details).} of $C_q(\X^i)$  allows us to further express this constraint set as
\begin{multline}
        \label{eq:range_Cq}
        \mc{S}_q(\X)\triangleq\range C_q(\X)= \\ \range X_1 \times
        \cdots \times \range X_{q-1} \times  \mathbb{R}^{I_{M_q}\times Q} \\\times \range X_{q+1} \times \cdots \times \range X_K.   
\end{multline}
%\begin{equation}
%    \begin{split}
%        \min_{\X} \; &\varphi(\X^T\y) + \gamma(\A\X)\\
%        \st & \forall k\in\mc{K},\\
%        &\forall j\in\mc{J}_I^k,\; \eta_{j}(\X_k^T\y_k) \leq 0,\\
%        &  \forall j\in\mc{J}_E^k, \;\eta_{j}(\X_k^T\y_k) = 0\\
%        & \X_k \in \{\bm{U} \;|\; \exists {\lX_k} : \X_k=\X_k^i\lX_k \} \text{
%        if } k\neq q,
%    \end{split}
%\end{equation}
The new constraint thus simply restricts each block $\X_k$ of the optimization variable $\X$ to stay within the range of the corresponding
block $\X_k^{i}$ of $\X^i$ (at the previous iteration), except for the block associated with the
updating node $q$, which can move freely within the original constraint set. Solving
this equivalent problem and updating $\X^{i+1}$ with its solution is effectively
equivalent to performing the \emph{Local Solution} and \emph{Parameters Update}
steps described earlier in Section \ref{sec:algorithm_description}. 

\subsection{Notation and Proof Outline}
%\begin{equation}
%\label{eq:constraints_set}
%\begin{split}
%    \mc{X} \triangleq \{\X\in\mathbb{R}^{M\times Q} \;|\; & \forall k\in\mc{K},\\
%    \forall & j\in\mc{J}_I^k,\; \eta_{j}(\X_k^T\y_k, \X_k^T\C_{j,k}) \leq 0,\\
%\forall & j\in\mc{J}_E^k, \;\eta_{j}(\X_k^T\y_k, \X_k^T\C_{j,k}) = 0\}
%\end{split}
%\end{equation}
%describing the feasible set. % as a parametric set depending on the data $\y$ and $\C$. We use $\mc{X}$ without any argument to denote the global feasible set $\mc{X}(\y,D)$. 

%Using this notation, we can also express the local problems solved by each node in terms of the global variable $\X$ instead of the local variable $\lX$ (which is inconvenient to work with as it has a different interpretation at each updating node). 
We can summarize the local problem at any node $q$ by a single set-valued map
\begin{equation}
    \label{eq:map_def}
    F_q(\X) \triangleq \argmin_{U\in \mc{X} \cap \mc{S}_q(\X)} \; L(U),
\end{equation}
such that \eqref{eq:local_problem_param} is equivalent to
\begin{equation}
\label{eq:alg_summar}
\X^{i+1} \in F_{q^i}(\X^i).
\end{equation}
We will show the convergence of the algorithm by studying the properties of the map
\eqref{eq:map_def}. We will first show that the successive
application of the map converges to the set of its fixed points, defined as
\begin{equation}
        \label{eq:fixed_points}
        \{ \X \;|\; \forall k\in\mc{K}, \; \X\in F_k(\X) \}.
\end{equation}
We will then show
that the fixed points are solutions, or at least stationary points, of the original problem
\eqref{eq:generic_problem}. In order to obtain the most general result, we do
not assume any particular order for the sequence of updating nodes $q^i$,
but simply that each node is selected infinitely many times. Formally, this can
be expressed as 
\begin{equation}
        \label{eq:q_i}
        \textrm{Acc } \seq{q}{\N} = \mc{K}.
\end{equation}
where $\textrm{Acc }\cdot$ denotes the set of accumulation points of a
sequence\footnote{An accumulation point of a sequence is defined as a point that has infinitely many elements of the sequence in a fixed neighborhood around itself, no matter how small the neighborhood. A converging sequence has a unique accumulation point.}. %This assumption is not restrictive, as it is readily satisfied by many update rules, including the cyclic update rule and the random update rule \cite{shi2016primer} (with an appropriate choice of distribution, such that every node is assigned a non-zero probability of being selected).

\subsection{Subsequential Convergence (in Fully-Connected Networks)}
\label{sec:conv}

This first section contains a proof that the NS-DASF algorithm converges to the set of its fixed points, i.e. for any distance $\delta$, we can find an index $T$, such that for any $i > T$ there is some fixed point $\X_F$ of NS-DASF such that $\norm{X_F-X^i}_F< \delta$.

This first part of the proof is organized as follows.
\begin{enumerate}
        \item We prove that the procedure results in a monotonic decrease of the objective, and a sequence of feasible points.
        \item We prove that the optimal function value of the local problem is an upper semicontinuous function of $\X^i$.
        \item Based on the above two results, we show that if we select a subsequence of $(\X^i)_{i\in\N}$ over which the updating node $q$ remains the same (i.e. $q^i=q$ for any $i$ in the subsequence), the accumulation points of such a sequence are fixed points of $F_q$.
        \item We finally show that this must also be true for every node $q$.
\end{enumerate}

\begin{propbox}
\begin{proposition}[Monotonic decrease]
    \label{prop:monotonic_decrease}
Let $\seq{\X}{\N}$ be a sequence of iterates generated by \eqref{eq:alg_summar}. Then, the sequence of objective function values
$(L(\X^i))_{i\in\N}$ is monotonically decreasing. In addition, all the points in $\seq{\X}{\N}$ are feasible, i.e. $X^i\in\mc{X}$.
\end{proposition}
\end{propbox}
\begin{proof}
        As by the definition of the algorithm \eqref{eq:alg_summar} $\X^{i+1}$ is a solution of \eqref{eq:map_def}, it must be feasible (i.e. $\X^{i}\in\mc{X}$ for any $i$).   By the definition \eqref{eq:range_Cq} of $\mc{S}_q$, we have that $\X\in\mc{S}_q(X)$ for every $\X$ (as each block of $\X$ is trivially in its own range).  Combining these two facts yields that
        $\X^i\in\mc{X}\cap\mc{S}_{q}(\X^i)$, i.e. it is feasible for the local problem \eqref{eq:local_problem_param}.
        As $\X^{i+1}\in F_q(\X^i)$, it is the solution of \eqref{eq:local_problem_param}, and therefore it must be that $L(\X^{i+1})\leq L(\X)$ for every $\X$ in $\mc{X}\cap\mc{S}_{q}(\X^i)$. Since we just showed that $\X^i$ is also in this set, we find that $L(\X^{i+1})\leq L(\X^i)$.
\end{proof}

The remainder of the convergence analysis relies on three technical assumptions, satisfied by a broad class
of problems.

\begin{assbox}
\begin{assumption}[Continuity, compactness and CCP regularizer]
        \label{ass:cont}
        The function $L:\mc{X}\mapsto \mathbb{R}$ is continuous and has compact
        sublevel sets. In addition, $\gamma$ is closed, convex and proper (CCP).
\end{assumption}
\end{assbox}
The monotonic decrease property from Proposition \ref{prop:monotonic_decrease} implies that the compactness of all sublevel sets can be relaxed to the compactness of at least the sublevel set of $L(\X^0)$ in $\mc{X}$, where $\X^0$ is the initialization point of the algorithm.

\begin{assbox}
\begin{assumption}[Similar solutions]
        \label{ass:sub_sols}
The solutions of $\mathbb{P}(\cdot,\cdot,\cdot,\cdot)$ are unique up to a full-rank
transformation, i.e. they share the same column space.
\end{assumption}
\end{assbox}
This is trivially satisfied by strictly convex problems, or problems that can be cast as subspace problems, i.e., optimization problems where the solution set is a linear subspace\footnote{Of which principal component analysis is a notable example.}. This assumption ensures
that in the case of local problems with multiple solutions, all the solutions yield the same local problem at the next iteration, that is the value of
$\mc{S}_q(X^{i+1})$ is independent of any particular choice of local solution $\lX^\star$. Indeed, one may notice from the definition of $\mc{S}_q$ in \eqref{eq:range_Cq} that for any full-rank matrix $R$, $\mc{S}_q(\X R)=\mc{S}_q(\X)$.%as from the definition \eqref{eq:def_Cq} of $C_q(X)$,  $C_q(\X R)$ and $C_q(\X)$ have the same range and hence
%\begin{align*}
%        \X \in \{U\in\mathbb{R}^{M\times Q} &\;|\; \exists {\lX} : U={C}_q(\X)\lX \}=\mc{S}_q(X)\\
%&\Leftrightarrow \X \in \range({C}_q (\X)) \\
%&\Leftrightarrow \X \in \range({C}_q (\X R))\\
%&\Leftrightarrow R\X \in \mc{S}_q(X R).
%\end{align*}
%

The next assumption relies on a parametric description of the constraint set. We define the set-valued map
\begin{equation}
\label{eq:constraints_map}
\begin{split}
        \mc{X}(\y,\mc{D}) \triangleq \{\X\in\mathbb{R}^{M\times Q} \;|\; & \forall k\in\mc{K},\\
    \forall & j\in\mc{J}_I^k,\; \eta_{j}(\X_k^T\y_k, \X_k^T\C_{j,k}) \leq 0,\\
\forall & j\in\mc{J}_E^k, \;\eta_{j}(\X_k^T\y_k, \X_k^T\C_{j,k}) = 0\}
\end{split}
\end{equation}
which describes the feasible set $\mc{X}$ as a parametric set depending on the data $\y$ and $\mc{D}$. We will keep using $\mc{X}$ without any argument to denote the feasible set of the global problem \eqref{eq:generic_problem}. 
\begin{assbox}
\begin{assumption}[Feasible set continuity]
        \label{ass:lsc}
      %  For any $k\in\mc{K}$, the set-valued map $\mc{R}_k:\X\to
      %  \mc{X} \cap \mc{S}_k(\X)$ is both upper and lower
      %  semicontinuous
        The set-valued map $(\y,\mc{D})\mapsto \mathcal{X}(\y,\mc{D})$ is continuous, i.e., it is both upper and lower semicontinuous\footnote{This particular concept of set-continuity has different
        names depending on fields and authors. Authors in variational
        analysis typically use
        ``inner'' and ``outer'' semicontinuity
        \cite{rockafellar2009variational}, authors in mathematical economics use upper and lower
``hemicontinuity'' \cite{Charalambos2013}, and authors in set analysis use the
terminology we adopt in this paper \cite{aubin2009set}.}.
\end{assumption}
\end{assbox}

Set continuity can intuitively be understood as requiring that the minimum distance between any point in the set resulting from an infinetisimal change in the inputs of
$\mc{X}$, i.e., perturbations in the distribution of $\y$ or in the entries of $\mc{D}$, and the points of the original set can always be made arbitrarily small by choosing a sufficiently small input perturbation (i.e. the set grows and shrinks ``smoothly''). To further illustrate this property, we give some examples of continuous parametric constraint sets.

\paragraph{Linear constraints}
Linear constraints of the form
\begin{equation}
        \label{eq:linear_constraints}
        \mc{X}(R) = \{ \X \;|\; R \X = P \}
\end{equation}
vary continuously with $R$ on trajectories where $R$ has constant
rank. Indeed, the solution of this linear system can be expressed via the
pseudo-inverse of $R$ (used to express the projection on the null-space of
$R$), which is continuous on the set of matrices with
constant rank \cite{rakovcevic1997continuity,Forchini2005}. Intuitively, if the range of $R$ would suddenly lose a dimension,
then a new dimension would appear in the solution space of $R\X=P$ (corresponding to $R$'s
null-space having gained a dimension).

\paragraph{Orthogonality constraints}
Quadratic orthogonality constraints of the form
\begin{equation}
        \label{eq:quad_constraints}
        \mc{X}(\bm{y}) = \{ \X \;|\; \X^T\E{\y\y^T}\X ={I} \}
\end{equation} are continuous on trajectories where $\E{\y\y^T}$ has full
rank. A proof of this fact can be found in the supplementary materials.

\paragraph{Convex Inequality Constraints}
Convex inequality constraints of the form
\begin{equation}
        \label{eq:convex_ineq_constraints}
        \mc{X}(\bm{y}, \mc{D}) = \{ \X \;|\; G(\y,\X, \mc{D}) \preceq 0 \}
\end{equation}
are continuous if the function $G$ is convex and continuous in $\X$ for any $\mc{D}$ and distribution of 
$\bm{y}$, and if for any $\bm{y}$ and $\mc{D}$ there is a strictly feasible $\X$. A proof can be found in \cite{rockafellar2009variational}.

        Continuity of $\mathcal{X}$ also ensures the continuity of the map $\mc{R}_q: \X \mapsto \mc{X}\cap\mc{S}_{q}(\X)$, describing the local constraint set, as stated in the following lemma.

\begin{lembox}
\begin{lemma}[Continuity of Local Feasible Sets]
        \label{lem:Rq_cont}
        Let $(\y,\mc{D})\mapsto \mathcal{X}(\y,\mc{D})$ be a continuous map. Then the map $\mc{R}_q: \X \mapsto \mc{X}\cap\mc{S}_{q}(\X)$ is continuous.
\end{lemma}
\end{lembox}
\begin{proof}
        See supplementary materials.
\end{proof}

\begin{remark}
        In the two first constraint set examples, the continuity of the constraint set requires the data to have constant rank. Keeping in mind that 
        \begin{equation}
        \mc{X}\cap\mc{S}_{q}(\X^i)=\mc{X}(\ly^i,\widetilde{\mc{D}}^i),
        \end{equation}
        this constant rank assumption must also hold for the subblocks of $\X$, as if any $\X_k^i$ loses rank, the compressed $\ly_k$ will have dependent channels. When the continuity of the constraint set is only valid for full-rank data,  some additional mechanism must ensure that this condition is met. Slight algorithmic modifications to DASF have been proposed in \cite{musluoglu2022unified_p2} for the cases where this could not be guaranteed. Similar modifications can be applied to NS-DASF (we refer to \cite{musluoglu2022unified_p2} for further details). 
\end{remark}

We now turn our attention to the actual proof.
%\subsubsection{Convergence to a fixed point}
%This first part of the proof does not depend on the problem structure, and therefore applies to both smooth and non-smooth variants of DASF (i.e. the particular problem hidden behind the map $F_q$).
As the sublevel sets of $L$ are compact (Assumption \ref{ass:cont}), any sequence $\seq{\X}{\N}$ satisfying
the update rule \eqref{eq:alg_summar} has at least one accumulation point $\accX$
\cite[Theorem 3.6]{rudin1976principles}. In other words, there is some index set $\I\subseteq\N$ such that
$\seq{\X}{\I}$ converges to $\accX$. Because of \eqref{eq:q_i}, we can in
particular select $\I$ such that $\seq{q}{\I}$ is a constant subsequence with
value $q$. Our first objective is to show that the accumulation point $\accX$
associated with such a sequence is a fixed point of the map $F_q$. Before doing
so, we first need to show that the value function 
\begin{equation}
        \label{eq:value_function}
        m_q(\X) \triangleq \min_{U \in \mc{R}_q(\X)} L(U)
\end{equation} 
is upper semicontinuous (in the classical sense, not the set-analysis sense), as stated in the following lemma. 
\begin{lembox}
\begin{lemma}[Semicontinuity of the Value Function]
        \label{lem:usc}
Under Assumptions \ref{ass:cont} and \ref{ass:lsc},
the value function $m_q:\mc{X}\to\mathbb{R}$ is upper semicontinuous.
\end{lemma}
\end{lembox}
\begin{proof}
        See Appendix \ref{apx:proof_usc}.
\end{proof}
We can now relate the accumulation points of $\seq{\X}{\N}$ to the fixed points
of the maps $F_q$.
\begin{lembox}
\begin{lemma}
        \label{prop:fixed_point_q}
        Let $\seq{\X}{\N}$ and $\seq{q}{\N}$ be sequences related by
        \eqref{eq:alg_summar} and let $\mc{I}\subseteq \N$ be such that the
        subsequence $(\X^i, q^i)_{i\in\I}$ converges to $(\accX, q)$. Then under Assumptions \ref{ass:cont} and \ref{ass:lsc}, $\accX$
        is a fixed point of $F_q$.
\end{lemma}
\end{lembox}
\begin{proof}
        Since $q^i$ is part of the finite set $\mc{K}$, its convergence implies
        that it eventually becomes constant. We can therefore proceed as if it
        was a constant sequence.
        By \eqref{eq:alg_summar} and $\eqref{eq:value_function}$, we have
        $L(\X^{i+1})=m_q(\X^i)$. As $L$ is continuous with compact sublevel sets,
        it attains its minimum value in $\mc{X}$ (extreme value theorem) \cite{Charalambos2013}, and the sequence
        $L(\X^{i+1})$ is therefore bounded below by $L(\X^\star)$ and, as a consequence of its monotonic decrease (Proposition \ref{prop:monotonic_decrease}), it must therefore converge to some
        value $\bar{L}$ \cite{rudin1976principles}. From the
        continuity of $L$, it must be that $\bar{L}=L(\accX)$. Since $L(\X^{i+1})$ converges to $L(\accX)$, and since $m_q(\X^i)=L(\X^{i+1})$, it must hold that $m_q(\X^i)$ also converges to $L(\accX)$. From Lemma \ref{lem:usc} (and in particular
        \eqref{eq:usc}), it must be that 
        \begin{equation}
                \label{eq:Lltmq}
                L(\accX)\leq m_q(\accX).
        \end{equation}
        Since $\accX$ is the
        accumulation point of a sequence $(\X^i)_i$ living in the closed\footnote{The closedness of $\mc{X}$ follows from the continuity of the $\eta_j$.} set $\mc{X}$, it must be that
        $\accX\in\mc{X}$ (by the definition of closedness), and since $\accX\in\mc{S}_q(\accX)$ by definition (see \eqref{eq:range_Cq}), we conclude that $\accX\in\mc{R}_q(\accX)=\mc{X}\cap\mc{S}_q(\accX)$.
        Hence, we have by definition of the value function,
        \begin{equation}
                m_q(\accX)\leq L(\accX).
        \end{equation}
        Combining this fact with \eqref{eq:Lltmq}, we conclude that
        \begin{equation}
                \label{eq:val_eq_l}
                m_q(\accX)=L(\accX),
        \end{equation}
and hence $\accX\in F_q(\accX)=\argmin_{U\in\mc{R}_q(\accX)}L(U)$.
\end{proof}

It now remains to show that accumulation points of the full sequence $\seq{\X}{\N}$ are fixed
points of the map $F_q$ for any $q$. 

\begin{propbox}
\begin{proposition}[Subsequential convergence]
        \label{prop:fixed_point}
        Let $(\X^i)_{i\in\mathbb{N}}$ be any sequence generated by \eqref{eq:alg_summar}.
        Then under Assumptions \ref{ass:cont}, \ref{ass:sub_sols} and
        \ref{ass:lsc}, any accumulation point of $(\X^i)_{i\in\mathbb{N}}$ is
        a fixed point of the map $F_q$ for any $q\in\mc{K}$.
\end{proposition}
\end{propbox}
\begin{proof}
         Let $\seq{\X}{\N}$ be a sequence satisfying \eqref{eq:alg_summar}, by Assumpion \ref{ass:cont},
         there must be an index set $\mc{I}\subseteq \N$ such that the subsequence $(\X^i,\X^{i+1}, q^i, q^{i+1})_{i\in\mc{I}}$ converges to some $(\accX,\accX^{+1}, q, q')$ (we also used the fact that the cartesian product of compact sets is compact \cite{Charalambos2013}). We first prove that $\accX^{+1}\in F_q(\accX)$. From Assumption \ref{ass:lsc} and Lemma \ref{lem:Rq_cont},
        we have\footnote{This is the definition of set upper semicontinuity.}
        that $\accX^{+1}\in\mc{R}_q(\accX)$. From the continuity of $L$ (Assumption \ref{ass:cont}), we know that
        \begin{equation}
                \label{eq:acc_same_L}
        L(\accX^{+1})=L(\accX)=\lim_{i\to\infty}
        L(\X^i),
\end{equation}
        and by Lemma \ref{prop:fixed_point_q}  (see \eqref{eq:val_eq_l})
        $$L(\accX^{+1})=L(\accX)=m_q(\accX),$$
        we find that $\accX^{+1}\in F_q(\accX)$ (as $\accX^{+1}$ is in the local feasible set and minimizes the local objective). We will use this result to show that $\accX^{+1}$ is also a fixed point of $F_q$ (we already know that $\accX$ is a fixed point from Lemma 3). 

        From Assumption \ref{ass:sub_sols} and since both $\accX^{+1}, \accX \in F_q(\accX)$, we
        have
        $\mc{S}_k(\accX)=\mc{S}_k(\accX^{+1})$ and thus 
        \begin{equation}
                \label{eq:both_fixed_points}
        F_k(\accX)=F_k(\accX^{+1})
\end{equation}
        for any $k$, because, by Assumption \ref{ass:sub_sols}, they share the same column space. As any  pair of successive accumulation points $(\accX, \accX^{+1})$ share the same column space, we can inductively deduce that any sequence of accumulation points $(\accX, \dots,\accX^{+Q})$ (for the right selection of $\mc{I}$) share the same column space. In addition, by \eqref{eq:acc_same_L}, we also have
        \begin{equation}
                L(\accX)=L(\accX^{+1})=\dots=L(\accX^{+Q}).
        \end{equation}
        Because we assumed that every node is selected infinitely many times, i.e. $ \textrm{Acc } \seq{q}{\N}=\mc{K}$, for any $i$, we can select $Q<\infty$ such that the sequence of accumulation points associated with $(\X^i, \dots,\X^{i+Q})_{i\in\mc{I}}$ is such that all nodes are selected at least once between iterations (see supplementary materials for details). For any $q$ we can therefore find some $\accX^{+a}$, with $a\leq Q$, such that $\mc{S}_q(\accX) = \mc{S}_q(\accX^{+a})$ and hence because $L(\accX)=L(\accX^{+a})$,
        \begin{equation}
                \accX \in F_q(\accX),
        \end{equation}
        where $q$ was selected arbitrarily in $\mc{K}$.
 We therefore find that $\accX$ is a fixed point
        for any $q$ and therefore a fixed point of the full algorithm.  
\end{proof}

%At this point, we have proven that the algorithm converges to its set of fixed
%points (as by the definition of an accumulation point, a sequence converges to the set of its accumulation points (see Lemma \ref{lem:conv_to_acc}), which we have just shown to also be fixed points of the algorithm). This does not imply full convergence, but merely that the algorithm
%eventually oscillates between its fixed points (i.e. subsequential convergence). As shown in
%\cite[Theorem 4]{musluoglu2022unified_p2}, convergence to a single point can be guaranteed in
%the case where the set of fixed points (reachable by the local solver) is finite. We elaborate on the subject of convergence to a single point at the end of Section \ref{sec:optimality_fixed}.%\wip{This is not really what we prove in [12]: we either assume that the global problem has a finite number of stationary points, OR that the local subproblems have a finite number of (reachable) solutions. We do not mention anything about fixed points (as this is hard to check). So this statement comes a bit too early I feel (you first need to establish that the fixed points are stationary points). 

%By the way, I would also properly formalize this as we did in [12]. Convergence to a single point is important, so it could be better highlighted.

%Finally, isn't this only true for the case where you select the closest solution in case there are multiple solutions? You did not mention anything like that I think?}

\subsection{Optimality of Fixed Points (in Fully-Connected Networks)}
\label{sec:optimality_fixed}
It now remains to show that the aforementioned fixed points the algorithm converges to are optimal in some sense. We first describe the optimality result for the case of fully-connected networks, before describing it for arbitrary network topologies. For ease of notation let  us define the functions
\begin{equation}
                f(\X)\triangleq\varphi(\X^T \bm{y}(t), \X^T\B) \text{ and  }g(\X)\triangleq\gamma(\X^T\A).
\end{equation}
In the non-smooth case, a stationary point $\X^\circ$ is defined as a point satisfying the following
inclusion\footnote{A fairly complete and self-contained introduction to the
        notions of stationarity in
the non-smooth case is available in \cite{li2020understanding}.}:
\begin{equation}
        \label{eq:stationary_point}
        0\in \nabla f(\X^\circ) + \partial g(\X^\circ) + N_\mc{X}(\X^\circ)
\end{equation}
where $\nabla f(\X^\circ)$ is the gradient of $f$ at $\X^\circ$, $\partial g(\X^\circ)$
is the subdifferential of $g$ at $\X^\circ$ and $N_\mc{X}(\X^\circ)$ is the
normal cone to $\mc{X}$ at $\X^\circ$
\cite{Royset2021,rockafellar2009variational}. The sum of sets must be understood as a Minkowski sum\footnote{$\mc{A} + \mc{B}\triangleq \{a+b\;|\; a\in\mc{A},\;b\in\mc{B}\}$.}. Intuitively, the  normal cone to a
set at a particular point can be understood as the set of directions having no component ``pointing inwards''
the set (it therefore only contains 0 at a point in the set's interior). The subdifferential is the set of slopes of all the linear
underapproximators tangent to the graph of $g$ at a particular point. This inclusion translates the
fact that at a stationary point, any direction of improvement musts exit the
constraint set, and is known as Fermat's rule. In the smooth case, \eqref{eq:stationary_point} is strictly equivalent to the usual KKT conditions\cite{karush1939minima}, \cite{kuhnnonlinear}. 

In order to prove that the fixed points of the algorithm
satisfy \eqref{eq:stationary_point}, we will show that under the proper
assumption (or qualification, in optimization parlance), the local optimality at
each node (defined by \eqref{eq:map_def}) implies global stationarity (i.e. being a stationary point of the centralized problem \eqref{eq:generic_problem}).

We first treat the case of fully-connected networks before moving on to arbitrary topologies. %\wip{A proof for the smooth case with non-separable constraints is available in the supplementary materials.}

\subsubsection{Fully-Connected Networks}
\label{sec:convergence_fc}

We start with a technical lemma that ensures that the tangent cone of $\mc{R}_q(\X)$ is a subset of the sum of the tangent cones of $\mc{X}$ and $\mc{S}_q(\X)$.
%\wip{Make sure that the matrix definition is consistent with the vector definitions, add a note to make the link}
\begin{lembox}
\begin{lemma}
        \label{lem:sum_tangent_cones}
        Let $\X$ satisfy the following constraint qualification 
        \begin{equation}
                \label{eq:qual_fc}
                \forall U \in N_{\mc{X}}(\X), \quad
                C_k(X)^TU = 0\Rightarrow U = 0\quad \forall k
        \end{equation}
        Then for every $k$,
        \begin{equation*}
                N_{\mc{R}_k(\X)}(\X)\subseteq N_{\mc{X}}(\X)+N_{\mc{S}_k(\X)}(\X).
        \end{equation*}
\end{lemma}
\end{lembox}
\begin{proof}
        See Appendix \ref{apx:proof_sum_tangent_cones}.
\end{proof}
Later on, we will provide a more interpretable sufficient condition for the qualification \eqref{eq:qual_fc}, see Proposition \ref{prop:compressed_licq_arb}.

\begin{propbox}
\begin{proposition}[Optimality in Fully-Connected Networks]
        \label{prop:optimality-ns}
        Under Assumption \ref{ass:cont}, fixed points $\X^\circ$ of the algorithm $\eqref{eq:alg_summar}$
        satisfying the qualification \eqref{eq:qual_fc}
        are stationary
        points of problem \eqref{eq:generic_problem}. 
\end{proposition}
\end{propbox}
\begin{proof}
        Since $\X^\circ$ is a fixed point, it is a solution of the local problem \eqref{eq:local_problem_param} at any node $q$. This local optimality of $\X^\circ$ along with the fact that $\gamma$ and hence $g$ is CCP (Assumption \ref{ass:cont}) implies that \cite[Theorem 4.75]{Royset2021} for every $k$
        \begin{equation}
                \label{eq:local_optimality_2}
                0 \in \nabla f(\X^\circ) + \partial g(\X^\circ) + N_{\mc{R}_k(\X^\circ)}(\X^\circ).
        \end{equation}
        From \eqref{eq:local_optimality_2} and Lemma \ref{lem:sum_tangent_cones}, we find that the following must also hold:
        \begin{equation}
                \label{eq:local_optimality}
                0\in \nabla f(\X^\circ) + \partial g(\X^\circ) +
                N_\mc{X}(\X^\circ)+N_{\mc{S}_k(\X^\circ)}(\X^\circ).
        \end{equation}

        From \cite[Prop. 4.44 \& 6.43]{Royset2021}, the block separability of $g$ and $\mc{X}$ implies that
        \begin{multline}
                \label{eq:separable}
                \partial g(\X^\circ)+N_\mc{X}(\X^\circ)=(\partial
                g_1(\X^\circ_1)+N_{\mc{X}_1}(\X^\circ_1))\times\\ \dots\times (\partial
                g_K(\X^\circ_K)+N_{\mc{X}_K}(\X^\circ_K)).
        \end{multline}
        Therefore, as $N_{\mc{S}_k(\X^\circ)}$ also has a block structure, and in
        particular, the block
$[N_{\mc{S}_k(\X^\circ)}]_k =\{0\}$ (as the updating block is unconstrained, any
direction is feasible, i.e. $[\mc{S}_k]_k=\mathbb{R}^{M_k\times Q}$), we have that for any $k$,
        \begin{equation}
                -\nabla_k f(\X^\circ)\in\partial g_k(\X^\circ_k)+N_{\mc{X}_k}(\X^\circ_k)
        \end{equation}
        and therefore 
        \begin{equation}
                -\nabla f(\X^\circ)\in\partial g(\X^\circ)+N_{\mc{X}}(\X^\circ),
        \end{equation}
        which is equivalent to \eqref{eq:stationary_point}, hence
        completing the proof.

\end{proof}

%We also provide a result for smooth problems, see \cite{musluoglu2022unified_p2}
%for details. In order to provide an intuitive proof, we will consider a slightly stronger
%than necessary qualification, but easier to verify in practice. We refer to
%\cite{musluoglu2022unified_p1,musluoglu2022unified_p2} for a stronger result. Before we proceed, we define
%\begin{equation}
%        \bm{P}^\perp (\bm{X})=
%        \blkd(\bm{P}^\perp_{\X_1},\dots,\bm{P}^\perp_{\X_K}),
%\end{equation}
%where $\bm{P}^\perp_{\X_k}$ is the orthogonal projection onto the column space
%of a single block $\X_k$. The codomain of $\bm{P}^\perp$ is the smallest subspace
%orthogonal to all the subspaces $N_{\mc{S}_k(\X^\circ)}(\X)$ (or equivalently the subspace
%sum of all the $\mc{S}_k(\X)$). As a consequence, projecting the local
%optimiality conditions with $\bm{P}^\perp$ removes the dependency on $\mc{S}_k$
%at every node. 

We defer the statement of our main result for fully connected networks to Section \ref{sec:main_result}, after giving a more interpretable condition for the qualification of Lemma \ref{lem:sum_tangent_cones} to hold.

%\begin{figure}
%        \centering
%        \def\svgwidth{0.49\textwidth}
%        %\input{lsc_issue_alt.pdf_tex}
%        \caption{Proof of Proposition \ref{prop:convergence}}
%        \label{fig:proof_1}
%\end{figure}

\subsection{Extention to Arbitrary Network Topologies}
\label{sec:ext_Cconv_arb}

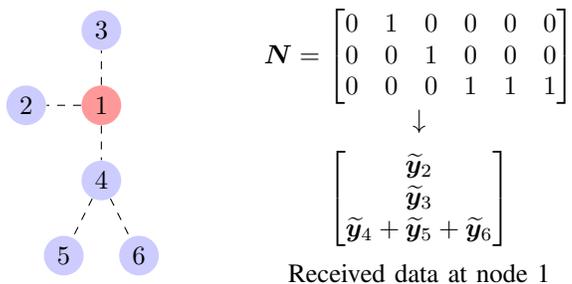
\begin{figure}[b]
        \begin{minipage}{0.24\textwidth}
        \centering
\tikzstyle{node}=[circle,draw=none,fill=blue!20,minimum size=1.5em,inner sep=0pt]
\tikzstyle{updating}=[circle,draw=none,fill=red!40,minimum size=1.5em,inner sep=0pt]
\begin{tikzpicture}
        \node[updating] (k1) at (0,0) {$1$};
        \node[node] (k2) at (-1,0) {$2$};
        \node[node] (k3) at (0,1) {$3$};
        \node[node] (k4) at (0,-1) {$4$};
        \node[node] (k5) at (-0.5,-2) {$5$};
        \node[node] (k6) at (+0.5,-2) {$6$};

        \path[dashed] (k1) edge (k2);
        \path[dashed] (k1) edge (k4);
        \path[dashed] (k1) edge (k3);
        \path[dashed] (k4) edge (k6);
        \path[dashed] (k4) edge (k5);

\end{tikzpicture}
\end{minipage}
\begin{minipage}{0.24\textwidth}
        \centering
        \begin{equation*}
        \bm{N}=\begin{bmatrix}
                0 & 1 & 0 & 0 & 0 & 0\\
                0 & 0 & 1 & 0 & 0 & 0\\
                0 & 0 & 0 & 1 & 1 & 1\\
        \end{bmatrix}
\end{equation*}
$$\downarrow$$
\begin{equation*}
        \begin{bmatrix}
                \ly_2\\
                \ly_3\\
                \ly_4+\ly_5+\ly_6
        \end{bmatrix}
\end{equation*}
Received data at node 1
\end{minipage}
\caption{An example aggregation matrix and received data for a tree rooted at
node 1.}
\label{fig:ex_network_and_matrix}
\end{figure}

In this subsection, we explain how all previous results, which were derived only for the case of a fully-connected network, can be extended to arbitrary topologies. It only requires a minor change in the description of the in-network fusion process. In the fully-connected case, this fusion process is mathematically described by the matrix $C_q$. For the case of arbitrary topologies, we have to modify this fusion matrix to describe the per-branch fusion within each of the trees constructed around each updating node $q$. Indeed, instead of receiving the compressed data $\cy^i_k=\X^{iT}_k
\y_k$
from all other nodes, the updating node $q$ will receive the linear combination
$T_q C_q(\X^i)^T \y$ where $T_q\in\mathbb{R}^{(M_q+|\mc{N}_q| Q)\times (K-1)Q+M_q}$ has the structure
\begin{equation}
        \label{eq:Tq}
        T_q= \blkd(I_{M_q},N\otimes I_Q)
\end{equation}
where $\otimes$ denotes the Kronecker product,  and $N$ is a topology-dependent
matrix encoding the aggregation procedure. More specifically, we define some
tree graph with the updating node as the root. The matrix $N$ groups the nodes based on the branches $\mc{B}_{nq}$ for $n\in\mc{N}_q$ (see Figure \ref{fig:ex_network_and_matrix} for an illustrative
example). It has
as many rows as the updating node has neighbors, and $K-1$ columns corresponding to all nodes with the column for node $q$ omitted. The element
$N_{i,j}$ is 1 if node $j$ is in the $i$-th branch, and 0
otherwise. The received
data $T_q C_q(\X^i)^T \y^i$ is then a block matrix, where the first block of $M_q$ rows correspond to the updating node's uncompressed data $\y_q$, and where each subsequent block of $Q$ rows
can be interpreted as the compressed data of a full branch (instead of a single node in the fully-connected setting). Note that in the case of fully-connected networks, $T_q=I$ for any $q$.

\subsubsection{Convergence}
The results of Sections \ref{sec:rel_loc_glob} to \ref{sec:conv} can be straightforwardly extended to arbitrary network topologies by replacing $C_q(\X)$ by $C_q(\X)T_q^T$ everywhere it appears in those sections, and thus update the definition of the local range constraint set $\mc{S}_q$ in \eqref{eq:range_Cq} with
\begin{equation}
        \mc{S}'_q(\X)\triangleq\range{C_q(\X)T_q^T}.
\end{equation}
$C_q(\X)T_q^T$ shares with $C_q$ the two properties that matters to the proof: its continuity and invariance to full-rank transformations of $\X$. The continuity ensures that the continuity of the local constraint set $\mc{X}\cap\mc{S}'_q(\cdot)$ in Lemma \ref{lem:Rq_cont} still applies to arbitrary topologies, and the invariance to full-rank transformation is a key property used in the proof of Proposition \ref{prop:fixed_point}. The proof of convergence is otherwise identical to the fully-connected case.

\subsubsection{Optimality}
Proposition \ref{prop:optimality-ns} must be slightly modified to ensure that the proper
qualification is met at each of the local problems, as described hereafter. %The qualification in the following proposition is the same as \cite[Condition 1b]{musluoglu2022unified_p2}.

\begin{propbox}
\begin{proposition}[Optimality in Arbitrary Network Topologies]
        \label{prop-ns-arb}
        Fixed points $\X^\circ$ of the procedure $\eqref{eq:alg_summar}$
        satisfying for every $k$ the qualification 
        \begin{equation}
                \label{eq:qual_arb}
                \forall U \in N_{\mc{X}}(\X^\circ), \quad
                T_kC_k(X^\circ)^TU = 0\Rightarrow U = 0
        \end{equation}
        are stationary
        points of the problem \eqref{eq:generic_problem}. 
\end{proposition}
\end{propbox}
\begin{proof}
        Let $\mc{R}'_k(\X)\triangleq \mc{X}(\C)\cap \mc{S}'_k(\X)$.
        The qualification is sufficient to ensure the inclusion \cite{rockafellar2009variational}:
 \begin{equation}
                N_{\mc{R}'_k(\X)}(\X)\subseteq N_{\mc{X}}(\X)+N_{\mc{S}'_k(\X)}(\X).
        \end{equation}
        The reasoning for the above inclusion is the same as Lemma \ref{lem:sum_tangent_cones}, but where we consider $T_kC_k(\X)^T$ instead of $C_k(\X)^T$.
        The rest of the proof is identical to the proof of Proposition \ref{prop:optimality-ns} with $\mc{R}_k$ replaced with $\mc{R}'_k$ and $\mc{S}_k$ replaced with $\mc{S}'_k$.
\end{proof}
Note that Proposition \ref{prop:optimality-ns} is a special case of Proposition \ref{prop-ns-arb} where $T_k=I$.

%For completeness, we also state the result for the smooth case, and refer the reader to \cite{musluoglu2022unified_p2} for the proof.
%\begin{proposition}[Optimality for smooth problems in arbirary topologies]
%        \label{prop:optimality_arb}
%        Let $L$ and $\mc{X}$ denote the objective and constraint set of Problem \eqref{eq:generic_problem_smooth}. Then fixed point of the procedure $\eqref{eq:alg_summar}$
%       satisfying the qualification 
%        \begin{equation*}
%                \forall U \in N_{\mc{X}}(\X^\circ), \quad
%                T_k^TC_k(X^\circ)^TU = 0\Rightarrow U = 0
%        \end{equation*}
%        are stationary
%        points of problem \eqref{eq:generic_problem_smooth}. 
%\end{proposition}

%\update{for block separability}

\subsection{Constraint Qualifications and an Upper Bound on the Number of Constraints}
The qualifications in Propositions \ref{prop:optimality-ns} and \ref{prop-ns-arb}
simply ensure that the local solutions satisfy the optimality conditions of the local
problems (see \cite{Royset2021} for examples failing to meet the
qualification). The following proposition gives a sufficient condition akin to the familiar linear independence constraint qualifications (LICQ) for the qualification to hold in the case of equality and inequality constraints in arbitrary topology networks.
In what follows, we define $\vartheta_j^k:\X_k\mapsto\eta_j(\X_k^T\y_k,\X_k^T\C_k)$.
\begin{propbox}
        \begin{proposition}[Constraint Qualification]
        \label{prop:compressed_licq_arb}
        Let $\mc{A}^k(\X)\triangleq \{j\in\mc{J}_I^k\;|\; \vartheta_j^k(\X_k)=0\}$ denote the set of active inequality constraints for node $k$ and 
        \[
                \label{eq:indep_pairwise}
                 \mc{D}_{nk}\triangleq  \left\{ \X^T_l\nabla\vartheta_j^l(\X_l)\;|\; j\in\mc{J}^l_E\cup\mc{A}^l(\X), l\in\mc{B}_{nk} \right\}.
           \]
        Then if the elements of $\mc{D}_{nk}$ are linearly independent matrices for every pair of neighboring nodes $n,k$, then
        it holds that 
        \begin{equation}
                \forall U \in N_{\mc{X}}(\X), \quad
                T_kC_k(X)^TU = 0\Rightarrow U = 0
        \end{equation}
        for any $k$.
\end{proposition}
\end{propbox}
\begin{proof}
        See Appendix \ref{apx:proof_compressed_licq_arb}.
\end{proof}
The independence of the elements of $\mc{D}_{nk}$ implies that the  ``compressed'' gradients $\X^T_l\nabla\vartheta_j^l(\X_l)$ of all the constraints, associated with all the nodes of a branch $\mc{B}_{nq}$ (for some updating node $q$), must be independent. In order to ensure that $\mc{D}_{nq}$ has independent elements, all the nodes in a single branch can (in total) have at most $Q^2$ constraints (the dimension of the compressed gradients). In the case of fully-connected networks, each branch contains a single node, and the bound can be relaxed to a maximum of $Q^2$ constraints per node. The difference with the similar bound found for the original smooth DASF algorithm \cite{musluoglu2022unified_p1, musluoglu2022unified_p2} can be attributed to the imposed separability of the constraints in the non-smooth problem \eqref{eq:generic_problem}. 

%Note that similarly to the fully-connected case, the qualification results in a
%upper limit on the allowable number of constraints. Indeed, for a given branch $\mc{B}_{nk}$, the dimension of the set \eqref{eq:indep_pairwise} is at most $Q^2$. As every node is eventually part of a branch, \eqref{eq:indep_pairwise} also implies that for any $k$
%\begin{equation}
%                \left\{ \X^T_k\nabla\vartheta_j^k(\X_k)\;|\; j\in\mc{J}^k_E\cup\mc{A}^k(\X), \right\}
%        \end{equation}
%        must have independent elements. The number of constraints per-node is therefore upper-bounded by $Q^2$, as in the fully connected case.
%At each iteration, the number of branches will be equal to $|\mc{N}_k|$, the problem can therefore have at most  $(|\mc{N}_k|+1)Q^2$ active constraints at each iteration. And therefore, we need to ensure that the total number of constraints $J$ is
%\begin{equation}
%        J \leq \min_k (|\mc{N}_k|+1)Q^2.
%\end{equation}
%We formalize this result in the following corollary.
%\begin{lembox}
%\begin{corollary}
%        If the number of per-nodes constraints is smaller than $Q^2$ and the total number of constraints is lower than $ \min_k (|\mc{N}_k|+1)Q^2$, then the qualification \eqref{eq:qual_arb} is satisfied with high probability.
%\end{corollary}
%\end{lembox}

\subsection{Main Result}
We finally summarize the convergence and optimality of NS-DASF with the following theorem:
\begin{theobox}
\begin{theorem}[Convergence and optimality]
        Let $(\X^i)_{i\in\N}$ be a sequence generated by Algorithm \ref{alg:fc} or \ref{alg:arb}. Assume that the qualification described in Proposition \ref{prop:compressed_licq_arb} holds at the accumulation points of $(\X^i)_{i\in\N}$, along with Assumptions \ref{ass:cont}-\ref{ass:lsc}. Then $(\X^i)_{i\in\N}$ converges to the set of stationary points of problem \eqref{eq:generic_problem}.
\end{theorem}
\end{theobox}
\begin{proof}
        By virtue of Proposition \ref{prop:fixed_point} (which can be extended to arbitrary topologies by considering $C_q(\X)T_q^T$ instead of $C_q(\X)$), the accumulation points of  $(\X^i)_{i\in\N}$ are fixed points of the algorithm, and the sequence therefore converges to the set of fixed points of the algorithm (see the supplementary materials for the proof that a sequence converges to the set of its accumulation points).  Finally, Proposition \ref{prop:compressed_licq_arb} ensures that the qualification \eqref{eq:qual_fc} is met, which in turns ensures by Proposition \ref{prop:optimality-ns} the stationarity of fixed points.
\end{proof}

\begin{remark}[Multiple Solutions]
        Although we only guarantee convergence to a set, we can in practice ensure convergence to a single point by various fixes. We can for example slighlty modify the problem by adding a regularizer ensuring a single solution, which ensures the existence of a single fixed point, hence making the set of fixed points a singleton. If modifying the problem is not possible, the algorithm can be sligthly modified, to select at each iteration the solution $\X^{i+1}$ minimizing the distance with the previous iterate $\X^i$. In addition, it can be shown that if the problem has a finite number of stationary points, then the algorithm always converges to a single point (see \cite[Theorem 4]{musluoglu2022unified_p2} for details).
\end{remark}

\label{sec:main_result}

\section{Numerical Experiments}
\label{sec:simulations}

This section provides some numerical results supporting the theoretical claims of Section \ref{sec:convergence}. We consider the problem of computing a sparse Wiener filter. The target filter is the solution of
\begin{equation}
    \label{eq:sim_problem}
   x^\star \triangleq \min_x \E{\norm{x^T \y(t) - \bm{d}(t)}^2_F}+\lambda \norm{x}_1,
\end{equation}
where $\bm{d}(t)$ is a single channel known target signal. Note that the regularization term can be written as $\norm{x^T A}_1$ with $A=I$, such that it fits \eqref{eq:generic_problem} (the $\ell_1$-norm is computed over a row vector here). We use a lowercase $x$ to emphasize that the filter has a single output channel (i.e. $Q=1$). $\lambda$ acts here as a meta-parameter allowing to tune the trade-off between solution accuracy and bandwidth (more nodes will eventually become inactive with a higher value of $\lambda$).
For the following experiments, we considered a fully-connected network where $M=K=10$. Each entry of a sample of $\y(t)$ is sampled from a zero-mean unit-variance gaussian distribution. $\bm{d}(t)$ is constructed as
\begin{equation}
    \bm{d}(t) = a^{ T}\bm{y}(t) + \bm{n}(t),
\end{equation}
where the entries in $\bm{n}(t)$ and $a$ are also sampled from a zero-mean unit-variance gaussian distribution.

The local problems associated with \eqref{eq:sim_problem} will be of the form
\begin{equation}
    \label{eq:sim_local_problem}
    \min_{\tilde{x}} \E{\norm{ \tilde{x}^T \ly(t) - \bm{d}(t)}^2_F}+\lambda \norm{ \tilde{x}^T\lA}_1.
\end{equation}

As $x\mapsto \norm{Ax}_1$ is not proximable, using proximal gradient descent \cite{beck2017first} for solving the local problems would be very inefficient. We use the Chambolle-Pock \cite{chambolle2011first} algorithm instead, as it was designed specifically for problems such as \eqref{eq:sim_local_problem}.

\begin{figure}
    \centering
    \includegraphics[width=0.49\textwidth]{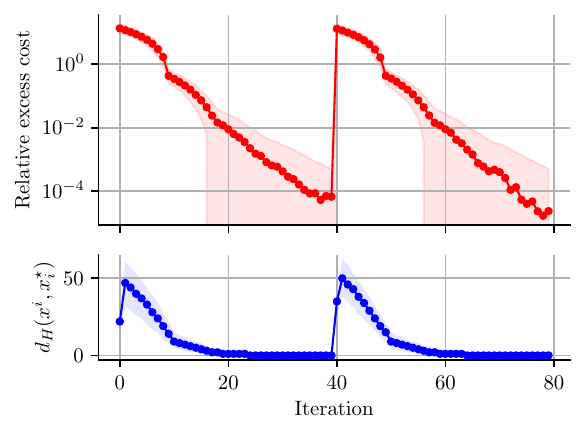}
\caption{Transient behavior of NS-DASF. The optimal solution changes at iteration 40. Solid curve depicts the median performance, the shaded area depicts the 5\%-95\% percentile region and were computed over 10000 Monte-Carlo runs.}
\label{fig:transient_sim}
\end{figure}

Figure \ref{fig:transient_sim} depicts the typical convergence behavior of NS-DASF applied to \eqref{eq:sim_problem}, i.e. error convergence when starting from a random filter. In order to better observe the adaptivity of the algorithm, the solution $x^\star$ was randomly changed at iteration 40. The top plot depicts the relative excess cost computed as 
\begin{equation}
    1 - \frac{L({x}^i)}{L({x}^\star)},
\end{equation}
and the bottom plot depicts the hamming distance between the set of active nodes, that is
\begin{equation}
    d_H(x_A,x_B) \triangleq H(|x_A| > 0, |x_B| > 0),
\end{equation}
where $H$ denotes the actual hamming distance, $|\cdot|$ and $>$ are the element-wise absolute value and ``greater than'' operators. 

NS-DASF applied to \eqref{eq:sim_problem} appears to exhibit a linear convergence rate. A first dip in excess cost can be observed after the tenth iteration,  as it is only then that each node has had the opportunity to freely update its corresponding block of $x^i$. The set of active channels is correctly identified after at most two full rounds of iterations (i.e. 20 iterations). Figure \ref{fig:transient_sim} depicts this same behavior when changing the solution to a new random point.

%\begin{figure}
%    \centering
%    \includegraphics[width=0.49\textwidth]{step.pdf}
%\caption{Convergence behavior of NS-DASF. Solid curve depicts the median performance, shaded area depics the 5\%-95\% percentile region and were computed over 10000 Monte-Carlo runs.}
%\label{fig:step_sim}
%\end{figure}

\begin{figure}
    \centering
\includegraphics[width=0.49\textwidth]{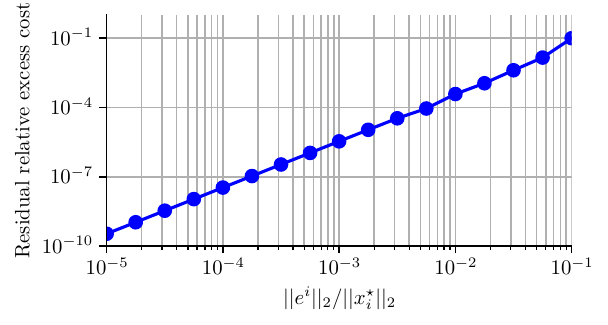}
\caption{Tracking performance of NS-DASF. Each data point corresponds to the mean relative excess cost at iteration 200 of 100 hundred Monte-Carlo runs.}
\label{fig:adaptive_sim}
\end{figure}

Figure \ref{fig:adaptive_sim} depicts the adaptive property of NS-DASF in a scenario where the solution changes over time for different rates of change. At each iteration, we apply a perturbation with zero-mean gaussian entries $e^i$ to the ground-truth solution $x^{i\star}$, with the ratio $\norm{e^i}_F/\norm{x^{i\star}}_F$ kept constant across iterations (this ratio reflects the amount of change in $x^\star$ across iterations, i.e., a higher ratio is more challenging). The residual excess cost is defined as 
\begin{equation}
    \lim_{i\to\infty} 1 - \frac{L({x}^i)}{L({x}^{i\star})},
\end{equation}
and is estimated in the simulation by computing the relative excess cost at iteration $i=200$.
%In order to limit the simulation time, the solver precision was capped to $10^{-8}$, resulting in an error floor ($10^{-8}$ in this case, as $L({x}^{i\star})\approx 1$). The dependence between the solution's rate of change and the residual excess cost appears to be linear.

\section{Discussion}
\label{sec:discussion}

In this paper, we have described an extension of the DASF algorithm for a particular class of non-smooth spatial filtering problems. We have provided theoritical results ensuring the convergence and optimality of this NS-DASF algorithm. Using Monte-Carlo simulations, we have demonstrated the algorithm's transient and stationary behavior when applied to a sparse form of the multichannel Wiener filtering problem. In particular, this shows that NS-DASF can be used to perform channel or node selection alongside a given filtering task, allowing energy savings by omitting the transmission from nodes with 0-norm filters.

\appendix
\label{sec:appendix}

\setlength\abovedisplayskip{0pt}
\setlength\belowdisplayskip{0pt}

\subsection{Proof of Lemma \ref{lem:usc}}
\label{apx:proof_usc}
\begin{proof}
Upper semicontinuity can be stated as follows: let $(\X^i,m_q(\X^i))_i$ be any
converging sequence  (note that this sequence is not necessarily an NS-DASF sequence). Let its limit be some $(\accX,\bar{m})$, then
 \begin{equation}
         \label{eq:usc}
         \bar{m} \leq m_q(\accX).
 \end{equation}
 We now attempt to prove this implication.
 From Lemma \ref{lem:Rq_cont}, for any $U\in \mc{R}_q(\accX)$, there exists\footnote{This is the definition of set lower semicontinuity
 \cite{rockafellar2009variational, Charalambos2013, aubin2009set}.} a sequence 
 $\seq{U}{\N}$ converging to $U$ and such that $U^i \in
 \mc{R}_q(\X^i)$. 
 From \eqref{eq:value_function} it follows that\footnote{The application of a function to a set denotes the image of that set under the function.} $m_q(\accX)\in L(\mc{R}_q(\accX))$, and we can therefore
 select some $U\in\mc{R}_q(\accX)$ such that $L(U)=m_q(\accX)$. By definition of $m_q$, the sequence $U^i\in\mc{R}_q(\X^i)$ converging to $U$ is such that $
 m_q(\X^i)\leq L(U^i)$.
 Taking the limit of both sides yields $\bar{m} \leq L(U)=m_q(\accX)$, completing the proof.
\end{proof}

\subsection{Proof of Lemma \ref{lem:sum_tangent_cones}}
\label{apx:proof_sum_tangent_cones}
\begin{proof}
Under the qualification, $\X$ satisfies for any $k$
      %  \begin{equation}
      %          N_{\mc{X}}(\X)\cap \nullsp(C(X)^T) \subseteq \{0\}
      %  \end{equation}
      %  As for every $k$, $\nullsp C_k(X)^T\subseteq \nullsp C(X)^T$, we have
        \begin{equation}
                \label{eq:no_null_int}
                N_{\mc{X}}(\X)\cap \nullsp C_k(X)^T \subseteq \{0\}.
        \end{equation}

        As $\nullsp C_k(X)^T$ is by definition the orthogonal complement of $\range C_k(\X)=\mc{S}_k(\X)$, we have $ \nullsp C_k(X)^T= N_{\mc{S}_k(\X)}(\X)$ (the normal cone to a linear subspace is
        its orthogonal complement \cite{Royset2021}), and hence \eqref{eq:no_null_int} implies that
        \begin{equation}
                \forall U\in N_{\mc{X}}(\X), \; U'\in N_{\mc{S}_k(\X)}(\X),\quad U'\neq U
        \end{equation}
        and $U\neq -U'$ (because $N_{\mc{S}_k(\X)}(\X)$ is a linear subspace)
        unless $U=0$,
        which implies \cite[Theorem 6.14]{rockafellar2009variational} that 
        \begin{equation}
                N_{\mc{R}_k(\X)}(\X)\subseteq N_{\mc{X}}(\X)+N_{\mc{S}_k(\X)}(\X).
        \end{equation}
\end{proof}

\subsection{Proof of Proposition \ref{prop:compressed_licq_arb}}
\label{apx:proof_compressed_licq_arb}
\begin{proof}
        We give a proof by contradiction.
        Assume
        that there is some $U\in N_{\mc{X}}(\X)$ such that $T_kC_k(\X)^TU = 0$ and $U\neq 0$. By construction, $U$ can be expressed as a non-null linear combination 
        \begin{equation}
                \sum_{m\in\mc{K}}\sum_{j\in\mc{J}_E^m\cup\mc{A}^m(\X)} \lambda_{j} \nabla\theta_j^m(\X)
\end{equation}
where $\theta_j^m: \X\mapsto \vartheta_j^m(\X_m)$.
        %\begin{equation}
           %     \label{eq:constraint_x_input}
%\end{equation}
        Because $T_kC_k(\X)^TU = 0$, it must be that for any $n\in\mc{N}_k$%$
        \begin{equation}
                \label{eq:lin_com_theta_2}
                \sum_{m\in\mc{K}}\sum_{j\in\mc{J}_E^m\cup\mc{A}^m(\X)}\lambda_{j}\sum_{l\in\mc{B}_{nk}} \X^T_l\nabla_l\theta_j^m(\X)=0.
        \end{equation}
        But we have for $l\neq m$ that 
        $\nabla_l\theta_j^m(\X)=0.$
        Therefore, from \eqref{eq:lin_com_theta_2}, it must be that for any $n$,
        \begin{equation}
                \sum_{m\in\mc{K}}\sum_{j\in\mc{J}_E^m\cup\mc{A}^m(\X)}\lambda_{j}\delta_{\mc{B}_{nk}}(m) \X^T_m\nabla\vartheta_j^m(\X_m)=0,
        \end{equation}
        where $\delta_{\mc{B}_{nk}}(m)$ is 1 if $m\in\mc{B}_{nk}$, $0$ otherwise.
        We can equivalently write 
        \begin{equation}
            \sum_{m\in\mc{B}_{nk}}    \sum_{j\in\mc{J}_E^m\cup\mc{A}^m(\X)}\lambda_{j} \X^T_m\nabla\vartheta_j^m(\X_m)=0,
        \end{equation}
        which contradicts the linear independence assumption \eqref{eq:indep_pairwise}, unless $\lambda_{j}=0$ for every $j\in \mc{J}_E^m\cup\mc{A}^m(\X)$ and $m\in\mc{B}_{nk}$.
\end{proof}

\bibliographystyle{IEEEtran}
\bibliography{main_biblio}
\end{document}